\documentclass[useAMS,usenatbib]{mn2e}
\usepackage{graphicx, txfonts, natbib}
\usepackage{epsf}
\usepackage{amssymb}
\usepackage{epstopdf}
\usepackage{longtable}
\usepackage{subfigure}
\usepackage{comment}
\usepackage[colorlinks=false,dvips]{hyperref}
\newcommand{\msun}{\mbox{M$_{\odot}$}}

\newcommand{\kms}{\mbox{$\rm{km}\,s^{-1}$}}

\DeclareMathAlphabet{\mathsc}{OT1}{cmr}{m}{sc}
\def\testbx{bx}%
\DeclareRobustCommand{\ion}[2]{%
\relax\ifmmode
\ifx\testbx\f@series
{\mathbf{#1\,\mathsc{#2}}}\else
{\mathrm{#1\,\mathsc{#2}}}\fi
\else\textup{#1\,{\mdseries\textsc{#2}}}%
\fi}

\newcommand{\Nai}{\ion{Na}{i}}

\newcommand{\Caii} {\ion{Ca}{ii}}

\newcommand{\Cii} {\ion{C}{ii}}

\newcommand{\SiII} {\ion{Si}{ii}}

\newcommand{\TIii} {\ion{Ti}{ii}}

\newcommand{\vsiii}{\ensuremath{v_{\ion{Si}{ii}}}}

\begin{document}
\title[PTF SN Ia spectra] {Exploring the spectral diversity of low-redshift Type Ia supernovae using the Palomar Transient Factory}
 \author[K. Maguire et al.]
 {K.~Maguire,$^1$\thanks{E-mail: kate.maguire@eso.org} M.~Sullivan,$^{2}$ Y.-C.~Pan,$^3$ A.~Gal-Yam,$^{4}$  I.~M.~Hook,$^{3,5}$ D.~A.~Howell,$^{6,7}$
 \newauthor    P.~E.~Nugent,$^{8,9}$  P.~Mazzali,$^{10,11,12}$ N.~Chotard,$^{13}$ K.~I.~Clubb,$^{8}$  A.~V.~Filippenko,$^8$    
 \newauthor  M.~M.~Kasliwal,$^{14}$  M.~T.~Kandrashoff,$^8$ D.~Poznanski,$^{15}$
    C.~M.~Saunders,$^{16,17}$  J.~M.~Silverman,$^{18}$ 
\newauthor  E.~Walker,$^{19}$   D.~Xu$^{20}$ \\ 
      $^1$European Southern Observatory for Astronomical Research in the Southern Hemisphere (ESO), Karl-Schwarzschild-Str. 2, 85748 Garching b. M\"unchen, Germany\\
      $^2$Physics \& Astronomy, University of Southampton, Southampton, Hampshire, SO17 1BJ, UK\\
      $^3$Department of Physics (Astrophysics), University of Oxford, DWB, Keble Road, Oxford OX1 3RH, UK\\
      $^4$Benoziyo Center for Astrophysics, Weizmann Institute of Science, 76100 Rehovot, Israel \\
      $^5$INAF -- Osservatorio Astronomico di Roma, via Frascati, 33, 00040 Monte Porzio Catone, Roma, Italy\\
      $^6$Las Cumbres Observatory Global Telescope Network, Goleta, CA 93117, USA \\
      $^7$Department of Physics, University of California, Santa Barbara, CA 93106-9530, USA \\
      $^8$Department of Astronomy, University of California, Berkeley, CA 94720-3411, USA \\
      $^{9}$Computational Cosmology Center, Lawrence Berkeley National Laboratory, 1 Cyclotron Road, Berkeley, CA 94720, USA \\
     $^{10}$Astrophysics Research Institute, Liverpool John Moores University, IC2, Liverpool Science Park, 146 Brownlow Hill, Liverpool L3 5RF, UK \\
     $^{11}$INAF -- Osservatorio Astronomico, vicolo dell'Osservatorio, 5, 35122 Padova, Italy\\
     $^{12}$Max-Planck Institut fu\"r Astrophysik, Karl-Schwarzschild-Str. 1, 85748 Garching b. M\"unchen, Germany\\
     $^{13}$Universit\'e de Lyon, Universit\'e Lyon 1, CNRS/IN2P3, Institut de Physique Nucl\'eaire de Lyon, 69622 Villeurbanne, France \\
      $^{14}$The Observatories, Carnegie Institution for Science, 813 Santa Barbara St, Pasadena, CA 91101, USA \\
      $^{15}$School of Physics and Astronomy, Tel Aviv University, Tel Aviv 69978, Israel \\
      $^{16}$Physics Division, Lawrence Berkeley National Laboratory, 1 Cyclotron Road, Berkeley, CA 94720, USA \\
      $^{17}$Department of Physics, University of California ,Berkeley, CA 94720-7300, USA \\
      $^{18}$Department of Astronomy, University of Texas, Austin, TX 78712-0259, USA \\
      $^{19}$Department of Physics, Yale University, New Haven, CT 06520-8121, USA\\
      $^{20}$Dark Cosmology Centre, Niels Bohr Institute, University of Copenhagen, Juliane Maries Vej 30, DK-2100 K{\o}benhavn {\O}, Denmark
 }
\maketitle

\begin{abstract}
We present an investigation of the optical spectra of 264 low-redshift ($z<0.2$) Type Ia supernovae (SNe Ia) discovered by the Palomar Transient Factory, an untargeted transient survey. 
We focus on velocity and pseudo-equivalent width measurements of the \SiII\ 4130, 5972, and 6355\,\AA\ lines, as well those of the \Caii\ near-infrared (NIR) triplet, up to +5\,days relative to the SN $B$-band maximum light. We find that a high-velocity component of the \Caii\ NIR triplet is needed to explain the spectrum in $\sim 95$ per cent of SNe Ia observed before $-5$\,days, decreasing to $\sim 80$ per cent at maximum.  The average velocity of the \Caii\ high-velocity component is $\sim 8500$ \kms\ higher than the photospheric component.   We confirm previous results that SNe Ia around maximum light with a larger contribution from the high-velocity component relative to the photospheric component in their \Caii\ NIR feature have, on average, broader light curves and lower \Caii\ NIR photospheric velocities. We find that these relations are driven by both a stronger high-velocity component and a weaker contribution from the photospheric \Caii\ NIR component in broader light curve SNe Ia. We identify the presence of \Cii\ in very early-time SN Ia spectra (before $-10$\,days), finding that $>$40 per cent of SNe Ia observed at these phases show signs of unburnt material in their spectra, and that \Cii\ features are more likely to be found in SNe Ia having narrower light curves.

\end{abstract}


\section{Introduction} \label{intro}

The use of Type Ia supernovae (SNe Ia) as cosmological probes is firmly established \citep[e.g.,][]{1998AJ....116.1009R,2007ApJ...659...98R,1999ApJ...517..565P,2009ApJS..185...32K,2011ApJ...737..102S,2012ApJ...746...85S,2013MNRAS.433.2240G,2013arXiv1310.3828R,2014arXiv1401.4064B}. While there is reasonable consensus that the stars that explode as SNe Ia are CO white dwarfs (WDs) in binary systems, the nature of their companion stars is still controversial with a number of scenarios considered plausible. In the single-degenerate (SD) model, a nondegenerate companion star is the mass donor \citep{1973ApJ...186.1007W}, while in the double-degenerate (DD) model, a thermonuclear explosion results from the merger of two CO WDs \citep{1984ApJS...54..335I, 1984ApJ...277..355W}. The `double-detonation' scenario where the explosion is initiated by the detonation of a He layer on the WD surface, which sends a shock wave into the star resulting in a second detonation that unbinds the star, is experiencing a revival \citep{1990ApJ...361..244L,2014ApJ...785...61S}. There is also increasing evidence that there is more than one progenitor channel contributing to the SN Ia population, although the relative rates of the different channels remain unclear \citep{2010Natur.463..924G,2011Natur.480..348L,2011Natur.480..344N,2011Sci...333..856S,2012Sci...337..942D,2012Natur.481..164S,2013MNRAS.436..222M,2013arXiv1312.0628M,2013Sci...340..170W}.

A large diversity in the observed properties of SNe Ia has been identified, particularly from studies of their light curves and spectra at maximum light. Good progress has been made through the analyses of large spectroscopic datasets that have found strong differences in the presence (or absence) of certain spectral features, line strengths, and velocities within the samples \citep[e.g.,][]{ben05, 2006PASP..118..560B,2008AJ....135.1598M,blo12,2012MNRAS.425.1819S}. Correlations between spectroscopic properties and light-curve shape and colour, as well as between spectral features and host-galaxy properties \citep[e.g.,][]{2008A&A...477..717B,2012ApJ...748..127F,2012MNRAS.426.2359M}, have been identified at varying significance, dependent on the samples and the analysis techniques used. Some studies have specifically focussed on investigating trends between spectral features and luminosity, with the aim of decreasing the scatter in the SN Ia Hubble diagram \citep{2008A&A...477..717B,2009A&A...500L..17B,2011A&A...526A..81B,2011MNRAS.410.1262W,2012MNRAS.425.1889S}. Potential connections between the progenitor systems (detectable through signatures of circumstellar material) and observed properties such as light-curve shape, light-curve colour, and host-galaxy properties have also been made \citep{2012ApJ...752..101F,2013MNRAS.436..222M}. 

Some of the most prominent features in SN Ia spectra around maximum light are produced by Si: the \SiII\ 6355 \AA, \SiII\ 5972 \AA, and \SiII\ 4130 \AA\ lines. The ratio of the strength of the \SiII\ 6355 \AA\ line to that of the \SiII\ 5972 \AA\ line is well known to correlate with SN light-curve width \citep{1995ApJ...455L.147N}. The equivalent widths of these two \SiII\ features and the velocity gradient of the \SiII\ 6355 \AA\ line can also be used as a diagnostic for identifying subclasses of SNe Ia \citep{ben05, 2006PASP..118..560B}. Recent work has highlighted the importance of studies of the \Caii\ H\&K and \Caii\ near-infrared (NIR) triplet lines in SN  Ia spectra at early times -- previous studies have found a trend of increasing velocity of the complex \Caii\ H\&K absorption feature with increasing light-curve width \citep{1994AJ....108.2233W,1995ApJ...447L..73F,2012MNRAS.426.2359M}. It was speculated by \cite{2012MNRAS.426.2359M} that the presence of high-velocity features of the \Caii\ lines may cause this correlation but other causes such as \SiII\ 3858 \AA\ line contamination could not be ruled out. \cite{2013MNRAS.435..273F} suggested that a contribution from \SiII\ 3858 \AA\ was the more likely cause of this correlation and that the strength of the \SiII\ 3858 \AA\ varies strongly with light-curve width.

To investigate the origin of this relationship (contribution of high-velocity \Caii\ features or \SiII\ 3858 \AA\ to the \Caii\ H\&K feature), \cite{2014MNRAS.437..338C} studied 58 SNe Ia with spectra covering the \Caii\ NIR triplet, since this feature is less contaminated by features of other elements and should provide a much cleaner dataset. Using a multi-component fit, they found that the strength of the \Caii\ NIR high-velocity features also correlates with the light-curve width, such that SNe Ia displaying broader light curves have higher ratios of high-velocity to photospheric-velocity \Caii\ NIR features. This suggests that the origin of the previously identified trends in the \Caii\ H\&K feature are not caused predominantly by \SiII\ contamination but instead by high-velocity components of the \Caii\ absorption features. \cite{2014MNRAS.437..338C} also showed that SNe Ia with high \SiII\ velocities ($>$\,12000 \kms) tend to have weak \Caii\ NIR high-velocity features. \cite{2013Sci...340..170W} have identified an environmental dependence on \SiII\ velocities -- SNe Ia with higher \SiII\ velocities ($>$\,12000 \kms) preferentially occur in the centre of their host galaxies.

 \begin{figure*}
\includegraphics[width=18cm]{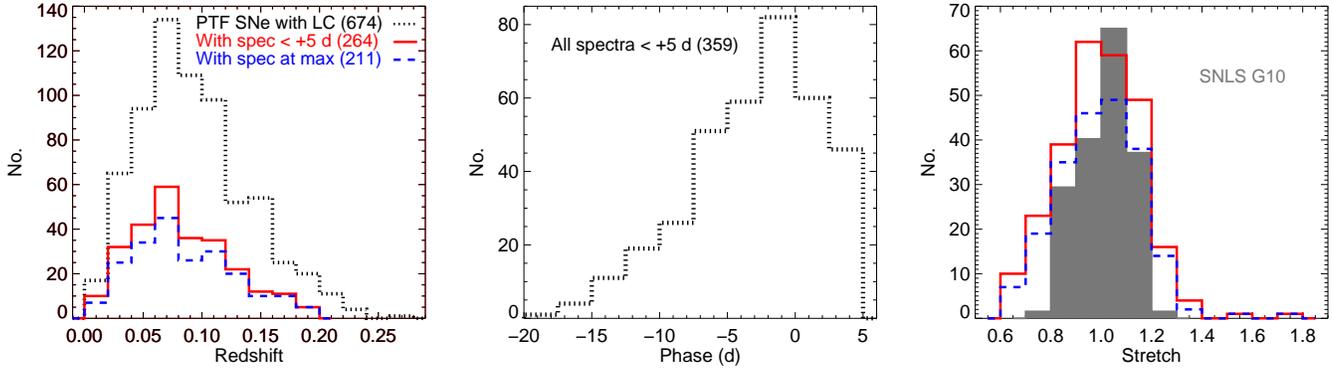}
\caption{\textit{Left panel:} Distribution of the redshift for all the PTF SNe Ia with measured light-curve parameters (black, dotted line), all SNe Ia with a spectrum at $<5$ d (red, solid line) and all those with a spectrum at `maximum light' ($-$5 to +5 d; blue, dashed line). The legend shows the numbers of SNe included in the different histograms. \textit{Middle panel}: The phase distribution of all spectra at $<5$ d. \textit{Right panel:} Stretch distribution for SNe Ia with a spectrum at $<5$ d (red, solid line) and with a spectrum in the `maximum light' sample (blue, dashed line). The SNLS stretch distribution of \protect \cite{2010A&A...523A...7G} at $z < 0.6$ is plotted as a grey, solid histogram.}
\label{dist}
\end{figure*}

The origin of the high-velocity features of \Caii\ is unclear but only small amounts of Ca are needed to produce strong \Caii\ features. Their high velocities (a few thousand \kms\ greater than the photospheric component) indicate that they are caused by material far out in the ejecta. High-velocity features are best identified at early times owing to the contrast between the photospheric and high-velocity feature velocities but this difference decreases with time, and without a multiple-component fit are difficult to distinguish by eye. These features were first discussed by \cite{2004ApJ...607..391G} and a number of suggestions for their origin have been made, including density and/or abundance enhancements at high velocity, either from circumstellar material (CSM) or intrinsic to the SN \citep{2005ApJ...623L..37M,blo12}. It has been shown that a small mass of H added to the outer layers of the ejecta, with reasonably high density at high velocities \cite[$\sim$4 times higher density at velocities $>$ 20,000 \kms\ than the standard W7 deflagration model of][]{1984ApJ...286..644N}, can lead to a substantial increase in recombination due to increased electron density \citep{2005ApJ...623L..37M,2007A&A...475..585A,2008ApJ...677..448T}. This increases the opacity of the strongest lines of singly ionised elements (in particular \Caii\ H\&K, \Caii\ NIR and then \SiII\ 6355 \AA) at high velocities, leading to the forming of high-velocity features. The \Caii\ fraction is increased by a factor of 5--10 compared to no H being added  \citep{2008ApJ...677..448T}. However, the link between the presence of these features and the progenitor configuration of SNe Ia remains elusive.

Identifying \Cii\ features in the early-time spectra of SNe Ia is of importance since \Cii\ traces the presence of unburnt material, providing a link to explosion mechanisms. Although O can also be used as a tracer of unburnt material, it too has a contribution from C burning.  Many searches for \Cii\ in SN Ia spectral samples have been performed, finding that $\sim 30$ per cent of SN Ia spectra obtained at epochs before $-$5 d with respect to maximum show signs of \Cii\ \citep{2011ApJ...743...27T, blo12,2012ApJ...745...74F,2012ApJ...752L..26P,2012MNRAS.425.1917S}. 

Different explosion mechanisms predict different amounts of unburnt material in early-time SN Ia spectra. Pure detonation models appear unviable since they predict too few intermediate-mass elements, and very little unburned material compared to observations \citep{1969Ap&SS...5..180A}. Pure deflagration models predict too much unburnt material in their ejecta, as well as the presence of O lines at nebular phases owing to mixing of the ejecta caused by turbulent burning \citep{2005A&A...437..983K,2005A&A...432..969R,2007A&A...464..683R}. However, this has only been observed in one subluminous SN Ia to date \citep{2013ApJ...775L..43T}. For the majority of SNe Ia, delayed-detonation models hold the most promise for explaining the observations, with the prediction that all but the outer layers of material will be burnt in the explosion \citep[e.g.,][]{1991A&A...245..114K,1995ApJ...444..831H,2010ApJ...712..624M}. Detections of \Cii\ signatures in early-time SN Ia spectra have allowed constraints to be placed on the amount and velocity structure of this outer, unburnt material. 
For example, \citet{2012ApJ...745...74F} showed that velocity of the \Cii\ 6580 \AA\ feature extends as low as $\sim 11000$ \kms\ (measured from the absorption minimum), suggestive of mixing of unburnt material to velocities similar to those of the \SiII\ lines.

Here we present optical spectra of SNe Ia obtained as part of the Palomar Transient Factory (PTF) collaboration. The PTF was an untargeted optical transient survey operating at the Samuel Oschin 48-in telescope (P48) at the Palomar Observatory, U.S. from 2009 to 2012 \citep{2009PASP..121.1395L,2009PASP..121.1334R}. Its aim was to discover transient events with timescales on the order of hours to years. It located 1249 low-\textit{z} spectroscopically confirmed SNe Ia, without biases in terms of host-galaxy properties (i.e., massive galaxies were not specifically targeted).  

 In this paper, we focus on the analysis of the spectroscopic features of these SNe Ia and the link to their photometric properties, specifically the SN light-curve width. This paper is complemented by Pan et al. (submitted), which details the connection between spectral properties at maximum light and SN Ia host-galaxy properties. Section \ref{obs_data} describes the sample selection, spectroscopic and photometric observations, as well as the data reduction and analysis techniques employed. Section \ref{analysis} presents the measurements made of the spectral features (velocities, equivalent widths, presence of high-velocity components) as a function of phase and light-curve width, as well as an analysis of high-velocity features and the presence of C in early-time spectra. The use of PTF SN Ia spectral measurements as luminosity indicators is also described. A discussion of the results of the previous sections are presented in Section \ref{discussion}, while Section \ref{conclusions} summarises the conclusions of this work. Throughout this paper, we assume a Hubble constant H$_0=70$\,km\,s$^{-1}$\,Mpc$^{-1}$.

\section{Observations and Data Reduction}
\label{obs_data}
 
We present spectroscopic classification and follow-up data for SNe Ia obtained as part of the PTF collaboration. SN candidates were discovered in P48 \textit{g}- and \textit{R}-band images using image-subtraction techniques. The SN candidates were ranked using machine-learning software \citep{2012PASP..124.1175B}, and visually confirmed by PTF members or citizen scientists via the  `Galaxy Zoo: Supernova project' \citep{2011MNRAS.412.1309S}. The best SN candidates were then classified \citep[for a review of SN types, see][]{1997ARA&A..35..309F} and monitored spectroscopically using a variety of optical telescopes. Photometry was obtained at the P48, as well as the 2-m robotic Liverpool Telescope \citep[LT;][]{2004SPIE.5489..679S} and the Las Cumbres Observatory Global Telescope (LCOGT) Faulkes Telescope North (FTN). In this paper, we focus on the spectroscopic sample analysis and use only derived quantities from the light curves (light-curve width, time of \textit{B}-band maximum light). The fully calibrated PTF SN Ia light curves will be presented in a future paper.

\subsection{Sample selection}
\label{samp_select}
We select SNe Ia for this spectroscopic study using the following criteria: (i) that a spectral comparison to SN templates is most similar to a SN Ia, (ii) that their light-curve parameters can be measured -- specifically that there are enough data available so that a light-curve width (stretch, \textit{s}) can be measured with the light-curve fitter, SiFTO \citep{2008ApJ...681..482C} with an uncertainty $<0.15$, (iii) that a spectrum exists at a phase of $<+$5 d with respect to \textit{B}-band maximum, (iv) that a spectroscopic redshift is available for the SN host galaxy (see Section \ref{opt_spec}) and has a value of $z<0.2$, and (v) that the spectrum has a sufficiently high signal-to-noise ratio (S/N) that at least the velocity and pseudo-equivalent width (pEW) of the \SiII\ 6355 \AA\ can be measured. In this paper, we focus on `normal' SNe Ia and leave the analysis of the low-velocity unusual SNe (`2002es-like' and `2002cx-like' objects) found in PTF to White et al. (submitted). Only one SN (PTF10acdh) in our sample is further excluded after our initial cuts because of an overlap with the sample of White et al. (submitted). The spectrum of  PTF10acdh had a low S/N and gave a spectral fit consistent with a SN Ia. However, it had the lowest \SiII\ velocity of our sample, with a value of just 6800 \kms\ (measured from the absorption minimum) at $-$8 d with respect to maximum light, and it is most likely a  `2002es-like' SN.  

The number of SNe Ia remaining after each of the cuts is shown in Table \ref{tab:discard}. When these cuts have been made we are left with 359 spectra of 264 unique SNe Ia before +5 d, and 247 spectra of 211 unique events in a maximum-light sample ($-5$ to +5 d). Fig.~\ref{dist} shows the redshift, phase, and stretch distributions of the PTF SN Ia sample used here. For comparison, the stretch distribution from the SN Legacy Survey (SNLS) is also shown \citep{2010A&A...523A...7G}. The sample is limited to SNe Ia at $z<0.6$, since the SNLS sample is closer to being complete with this redshift cut than the full SNLS SN Ia dataset \citep{2010AJ....140..518P}.

\begin{table}
  \caption{Number of SNe Ia discarded by cuts on the light-curve quality and spectral information as detailed in Section \ref{samp_select}.}
 \label{tab:discard}
\begin{tabular}{@{}lccccccccccccccccccccccccccccc}
  \hline
  \hline
Cut& No. of SNe & Discarded SNe \\
\hline
Total &1249&--\\
LC with $s_{\rm err}<0.15$&674&575\\
With spectrum $<+5$ d &371&303\\
With spectroscopic $z<0.2$ &288&83\\
With \SiII\ 6355 \AA\ measurement$^1$ &264&24\\
\hline
      \hline
\end{tabular}
 \begin{flushleft}
$^1$ PTF10acdh is also excluded in this cut because of its SN 2002es-like properties as detailed in Section \ref{samp_select}. \\
\end{flushleft}
\end{table}

\subsection{Low-resolution optical spectroscopy}
\label{opt_spec}
\begin{figure}
\includegraphics[width=8.7cm]{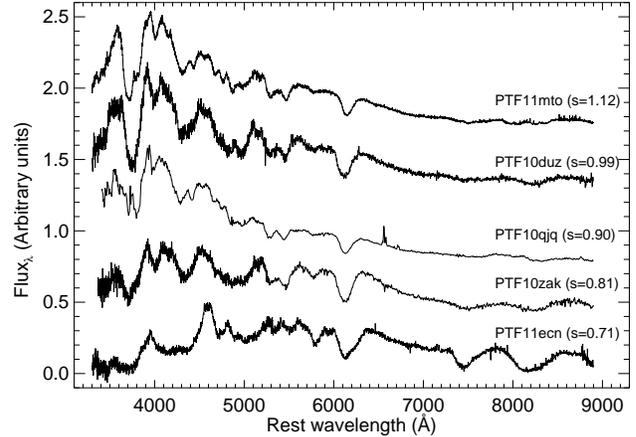}
\caption{Typical spectra from our maximum-light sample spanning the range of light-curve width values (stretch of $\sim$0.7--1.1). The SN name and its light-curve stretch is marked for each spectrum. }
\label{spec_ex}
\end{figure}

The optical spectra presented in this paper were obtained with 14 telescopes having 15 different instruments. The typical instruments and setups used for these observations are listed in Table \ref{tab:lowres-spec}, along with the number of spectra from each telescope entering our sample. The slit was generally aligned along the parallactic angle \citep[see][]{1982PASP...94..715F} to minimise the effects of atmospheric dispersion.  The spectra were reduced using custom pipelines for each of the telescopes based on standard spectral reduction procedures in \textsc{iraf} and \textsc{idl}. The two-dimensional spectra were bias and flat-field corrected before extraction. The extracted spectra were wavelength calibrated using arc-lamp exposures and instrumental response functions were obtained from observations of spectrophotometric standards to perform the flux calibration. As only velocity and equivalent-width measurements are presented in this paper, uncertainties in the absolute flux calibration do not significantly affect our results. The spectra will be publicly released via the WISeREP portal\footnote{http://www.weizmann.ac.il/astrophysics/wiserep/} \citep{2012PASP..124..668Y}. Example rest-frame spectra around maximum light for a range of light-curve widths are shown in Fig.~\ref{spec_ex}. 

The spectra were corrected to the rest frame using redshifts obtained with one of the following methods:  (i) from host-galaxy lines identified in the SN spectra, (ii) from NASA/IPAC Extragalactic Database\footnote{http://nedwww.ipac.caltech.edu/} (NED) galaxy spectra, (iii) from Sloan Digital Sky Survey Data Release 9 galaxy spectra \citep{2000AJ....120.1579Y}, or (iv) from host-galaxy spectra obtained after the SN had faded. Any SN Ia that did not have a redshift measured in this manner is excluded from our sample. This includes any with redshift measurements from template matching of the SN spectrum, since we are performing a velocity analysis and this could bias the results. However, a potential bias could be introduced by the removal of faint hosts, which are more likely not to have a host-galaxy spectrum. To quantify this potential bias, we compare the stretch distribution of the SNe Ia with and without a spectroscopic redshift using a Kolmogorov-Smirnov test, and find that their distributions are consistent with being drawn from the same parent population (p-value=0.83). 

\begin{figure}
\includegraphics[width=8.7cm]{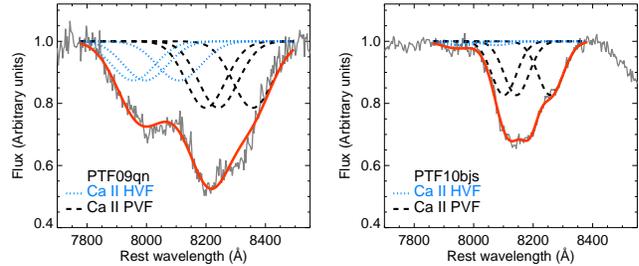}
\caption{Example spectral fits to the \Caii\ NIR spectral region for two SNe Ia in the maximum-light sample, PTF09qn and PTF10bjs. The combined fit to the normalised Ca II NIR region is shown as a red solid line, the PVF components as dashed black lines, and the HVF components as blue dotted lines. }
\label{spec_fit}
\end{figure}

\begin{table*}
  \caption{Spectral setups (telescope, instrument, spectral range) and the number of spectra per setup.}
 \label{tab:lowres-spec}
\begin{tabular}{@{}lccccccccccccccccccccccccccccc}
  \hline
  \hline
   Telescope+$^1$ &Typical $\lambda$&No. of spectra &No. of spectra\\
     instrument &range (\AA) & at $<$5 d& in max-light sample\\
\hline
\hline
  P200+DBSP&3200--10200&91&63 \\
  Keck1+LRIS&3200--11000&67&44 \\
  WHT+ACAM/ISIS&4950--9500, 3200--9500&61&48\\
  Lick+Kast&3400-10300&47&42\\
  UH88+SNIFS&3300--9700&28&13\\
   KPNO+RC&3400-8400&25&18\\
  Gemini-N+GMOS&3500--9700&25&9 \\
  VLT+XSH&3200--10200&3&2 \\
  Keck2+DEIMOS&4500--9600&3&2 \\
   TNG+DOLORES&3300-8200&3&1\\
 Gemini-S+GMOS&3500--9700&2&2\\
  APO+DIS&3200--9500&2&1\\
  Magellan+LDSS3&4000-10200&1&1\\
  Wise+FOSC&3900--8300&1&1\\
     \hline
\end{tabular}
 \begin{flushleft}
    $^1$Further information on the telescopes and instruments used:\\
    P200+DBSP = Palomar 200-inch (P200) with the Double Spectrograph \citep[DBSP;][]{1982PASP...94..586O}.\\
    Keck+DEIMOS/LRIS = Keck 10-m with the Deep Imaging Multi-Object Spectrograph \citep{2003SPIE.4841.1657F} and the Low-Resolution Imaging Spectrometer \citep{1995PASP..107..375O}. \\
    WHT+ACAM/ISIS = William Herschel Telescope (WHT) with the Advanced Camera (ACAM) and the Intermediate-dispersion Spectrograph and Imaging System (ISIS). \\
    Lick+Kast = Lick Observatory Shane 3-m telescope with the Kast double spectrograph \citep{miller_1993_lick}.\\
    UH88+SNIFS = University of Hawaii 2.2-m telescope (UH88) with the Supernova Integral Field Spectrograph \citep[SNIFS;][]{2004SPIE.5249..146L}  \\
    KPNO+RC = Kitt Peak National Observatory (KPNO) Mayall 4-m telescope with the RC spectrograph.  \\
    Gemini-N+GMOS = Gemini Telescope North (Gemini-N), Mauna Kea, Hawaii, US, with the Gemini Multi-Object Spectrograph \citep[GMOS;][]{2004PASP..116..425H}.\\
    VLT+XSH = Very Large Telescope (VLT) at Paranal, Chile with the XShooter (XSH) spectrograph \citep[XSH;][]{2006SPIE.6269E..98D,2011A&A...536A.105V}.\\
    TNG+DOLORES = 3.58m Telescopio Nazionale Galileo (TNG) with the Device Optimized for the LOw RESolution (DOLORES).\\
    Gemini-S+GMOS = Gemini Telescope South (Gemini-S), Cerro Pachon, Chile, with GMOS.\\
    APO+DIS = Apache Point Observatory (APO) with Dual Imaging Spectrograph (DIS). \\
    Magellan+LDSS3 = Magellan 6.5-m telescope with the Low Dispersion Survey Spectrograph-3 (LDSS3).\\
    Wise+FOSC = Wise Observatory 1-m telescope with the  Faint Object Spectrograph and Camera (FOSC).\\
\end{flushleft}
\end{table*}

\subsection{Optical photometry}
\label{opt_phot}

The optical photometry of the SN Ia sample comes from three telescopes: the P48, the LT and FTN. The LT is located at the Roque de Los Muchachos Observatory on La Palma, Spain, while FTN is located on Haleakala, Hawaii, U.S. The P48 provides $gR$-band data, which were reduced by the Infrared Processing and Analysis Center (IPAC)\footnote{http://www.ipac.caltech.edu/} pipeline \citep{2014arXiv1404.1953L} and photometrically calibrated \citep{2012PASP..124..854O}.  The LT data were obtained using both the RATCAM and IO:O optical imagers in $gri$ filters, similar to those used in the Sloan Digital Sky Survey. The FTN data were obtained with the Spectral Optical Imager in $gri$ filters.

Deep reference images of the SN field when the SN is not present (either pre-explosion or at $>$300 days post-explosion) using the same instrument and filters are necessary to remove host-galaxy contamination from the SN photometry. Since the PTF is a rolling search, P48 reference images are obtained pre-explosion by design. However, for the LT, a specific campaign was mounted to obtain these reference images after the SN had faded. For each SN image, the stacked reference images are registered, flux-scaled and subtracted off using a point-spread-function (PSF) matching routine.  The zeropoints of the LT images are calculated by calibrating directly to the SDSS photometric system using SDSS stars in the field of the SN or using calibration images (Landolt standards or SDSS Stripe 82 fields) taken before and after the SN fields on photometric nights. The FTN data have been reduced and calibrated in a similar manner to that described for the LT data.

For the calibration of LT data for the PTF SN Ia cosmology analysis, a more homogeneous and sophisticated calibration will be employed. In addition to SDSS Stripe 82 stellar fields, the updated cosmology standards stars of \cite{2013A&A...552A.124B} are being observed on photometric nights before and after deep reference images to allow more accurate calibration. However, this improved method has not yet been applied to the whole sample.

\subsection{Light-curve fitting}
\label{sec:light-curve-fitting}

The optical $gRri$-band light curves were analysed using the SiFTO
light-curve fitting code \citep{2008ApJ...681..482C}, which outputs values for the stretch, maximum \textit{B}-band magnitude, $B-V$ colour at maximum (where multiple bands are available), and time of maximum light for each SN. SiFTO uses a time series of spectral templates that are adjusted to match
the observed colours of the SN photometry at each epoch, while also
adjusting for Galactic extinction and redshift (i.e., the
K-correction). In this paper, we do not present the $B-V$ colours found with SiFTO but instead defer this to a future paper and focus on the connection between spectral features and light-curve width.

 \begin{figure*}
\includegraphics[width=16.5cm]{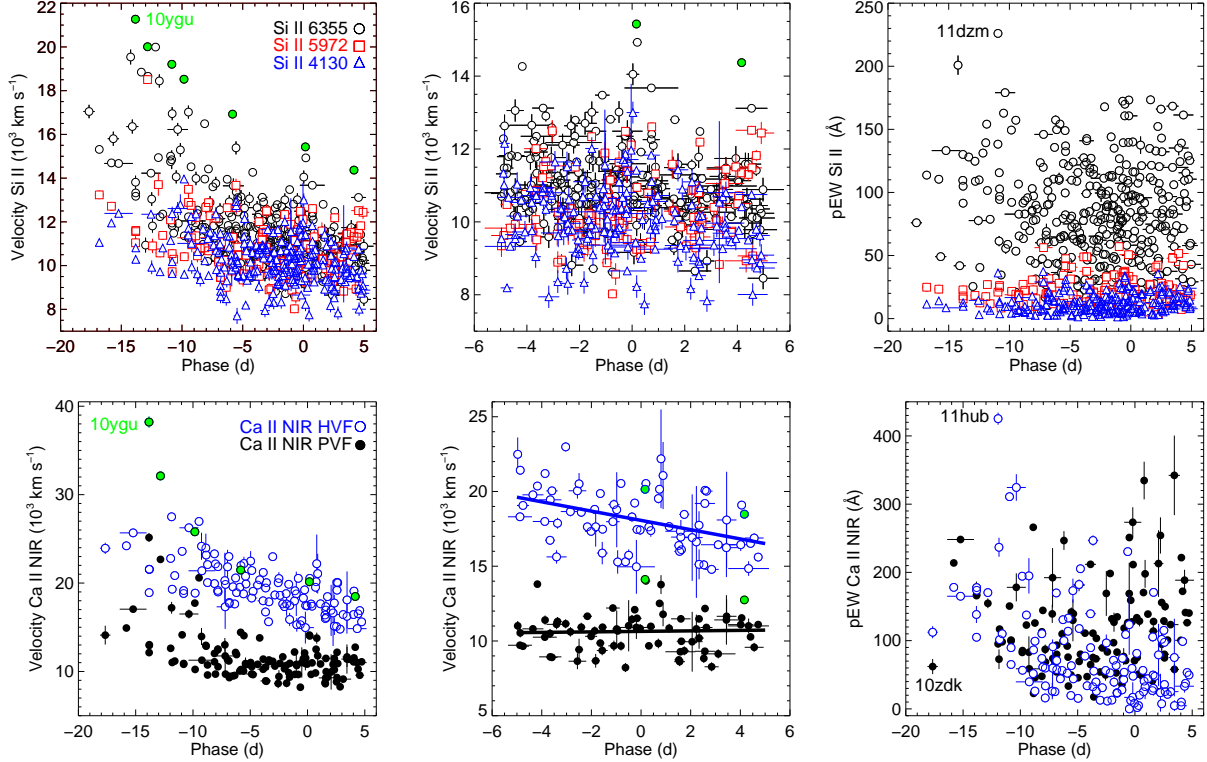}
\caption{\textit{Top-left panel:} The velocity of the \SiII\ 4130 \AA\ (blue, open triangles), \SiII\ 5972 \AA\ (red, open squares), and \SiII\ 6355 \AA\ features (black, open circles) as a function of phase. The \SiII\ 6355 \AA\ velocity of PTF10ygu \citep[SN 2010jn;][]{2013MNRAS.429.2228H} is marked with filled green circles. \textit{Top-middle panel:} Same plot as top-left panel but limited to the `maximum-light' sample ($-5$ to +5 d). The \SiII\ 6355 \AA\ velocity is marked for PTF10ygu with filled green circles. \textit{Top-right panel:} pEW of the \SiII\ 4130 \AA\ (blue, open triangles), \SiII\ 5972 \AA\ (red, open squares), and \SiII\ 6355 \AA\ (black, open circles)  lines as a function of phase with respect to \textit{B}-band maximum. \textit{Bottom-left panel:} The \Caii\ NIR triplet photospheric-velocity feature velocity (black, solid circles) and high-velocity feature velocity (blue, open circles) against phase in days since $\textit{B}$-band maximum. The \Caii\ NIR high-velocity component velocity is marked for PTF10ygu with filled green circles. \textit{Bottom-middle panel:} Same plot as bottom-left panel but limited to the `maximum-light' sample ($-5$ to +5 d). The best linear fits to the data are also shown, where the \Caii\ NIR HVF is seen to decrease more rapidly with time than the PVF. The \Caii\ NIR velocity of the high-velocity and photospheric-velocity feature for PTF10ygu is marked with filled green circles. \textit{Bottom-right panel:} pEW of the \Caii\ NIR triplet as a function of phase with respect to \textit{B}-band maximum. The pEW of both the \Caii\ NIR triplet photospheric-velocity component (black, filled circles) and high-velocity component (blue, open circles) are shown. }
\label{phase}
\end{figure*}

\subsection{Measuring line velocities and widths}
\label{fitting}

Different methods have been used in the past to fit SN Ia spectral features in optical spectra, including fitting single Gaussian profiles \citep[e.g.,][]{2012MNRAS.426.2359M}, spline fits \citep[e.g.,][]{2012MNRAS.425.1819S}, and applying smoothing techniques and then choosing the minimum value \citep[e.g.,][]{2006AJ....131.1648B,2011ApJ...742...89F}. Since we are particularly interested in the \Caii\ NIR triplet, which contains overlapping lines, we follow the multiple Gaussian fitting method of \cite{2013ApJ...770...29C,2014MNRAS.437..338C}. First, a pseudo-continuum is defined on either side of the absorption feature by visual inspection, and it is removed by fitting a line between the two points. The spectrum is then normalised at the feature position by removing the line fit. We perform the spectral fitting in velocity space, converting wavelengths to velocities using the relativistic Doppler formula.

Single or multiple Gaussian profiles can then be used depending on the line being fit. For the \SiII\ 6355 \AA\ line, a double Gaussian is used to account for the \SiII\  6347 \AA\ and \SiII\ 6371 \AA\ lines. To reduce the degeneracy of the fit, we assume an optically thick regime  \citep[typical of SN Ia atmospheres at early times; e.g.,][]{blo12} and that the lines tend to saturate. Therefore, we fix the strengths of the individual components of the doublet to be equal \cite[for a further discussion see][]{2013ApJ...770...29C}. We also fix the relative velocity difference between the double components, as well as force their widths to be the same. In a similar manner to the \SiII\ 6355 \AA\ feature, we measure the parameters of the \SiII\ 4130 \AA\ (doublet at 4131 and 4128 \AA) and \SiII\ 5972 \AA\ (doublet at 5979 and 5958 \AA) features. We constrain the relative strengths, velocity differences and widths of the doublet lines.

For the \Caii\ NIR triplet region, the fitting is more complicated. There are three separate components of the \Caii\ feature (8498, 8542, 8662 \AA). Since we wish to determine the potential contribution from `high-velocity' features, as well as photospheric-velocity features (PVFs), we fit six (three photospheric-velocity and three high-velocity) individual Gaussians simultaneously. We force each of the components of the \Caii\ NIR triplet PVF to have the same velocity and width, as well as constrain their velocities to be within 25 per cent of the \SiII\ 6355 \AA\ velocity; the latter is a looser constraint than that of \cite{2014MNRAS.437..338C} because we wish to investigate the connection between the \SiII\ 6355 \AA\ velocity and the \Caii\ PVF velocity. For the \Caii\ NIR high-velocity feature (HVF), we force the components to be at least 2000 \kms\ higher than the \SiII\ velocity (by varying this value and repeating the measurements, we find that 2000 \kms\ can be considered a conservative minimum). We also force the three components of the \Caii\ NIR HVF to have the same velocities, widths, and relative strengths, assuming an optically thick regime.  Example fits to the \Caii\ NIR region are shown in Fig.~\ref{spec_fit}.

Uncertainties in the velocities are estimated by propagating the uncertainty in the fitting with an additional one to account for uncertainties in the redshift of each SN. The quoted pEW uncertainties are those outputted from the fitting routine. 

\begin{figure*}
\includegraphics[width=18.cm]{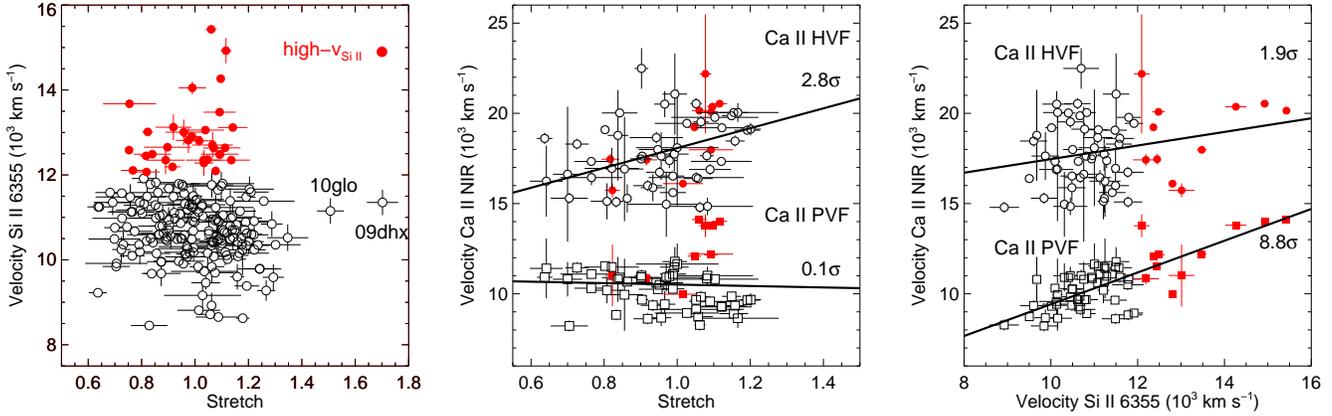}
\caption{\textit{Left panel:} \SiII\ 6355 \AA\ velocity against light-curve stretch for the maximum-light sample  (phase range $-5$ to +5 d). Two SNe Ia with very broad light curves (PTF10glo and PTF09dhx) are shown but are excluded from further analysis.  No correlation between the \SiII\ 6355 \AA\ velocity and the stretch was found. SNe Ia with \SiII\ 6355 \AA\ velocities above 12,000 \kms\ (`high-\vsiii') are shown with filled red circles, while the normal-velocity SNe Ia are shown as black open circles.  \textit{Middle panel:}  \Caii\ NIR velocities of the PVF and HVF against stretch. The open black circles and closed red squares represent SNe Ia with `normal' \SiII\ 6355 \AA\ velocities ($<$\,12,000 \kms) and `high' \SiII\ 6355 \AA\ velocities ($>$\,12,000 \kms), respectively. \textit{Right panel:} \Caii\ NIR velocities of the PVF and HVF against \SiII\ 6355 \AA\ velocities, colour-coded based on whether they have \SiII\ velocities above (red filled squares) or below (black open circles) 12,000 \kms. The solid black lines show the best linear fits to the data with the significance of the trends also shown.}
\label{col_str}
\end{figure*}

\section{Analysis}
\label{analysis}

In this section, we present the results of our spectroscopic analysis: (i) the properties of the spectral features as a function of the phase, (ii) the properties of the spectral features as a function of their light-curve width, (iii) the presence of high-velocity features and their relation to other properties, (iv) the breakdown of the sample into the spectroscopic subclasses of \cite{2006PASP..118..560B} and \cite{2009ApJ...699L.139W}, (v) the use of spectral lines as luminosity indicators, and (vi) the search for unburnt material in the form of \Cii.

\subsection{Evolution of spectral features with phase}
 
Here we detail the spectral velocities and pEW measured using a Gaussian fitting technique (discussed in Section \ref{fitting}) as a function of phase. We focus in this analysis on general trends with phase for the SNe Ia in our sample, instead of on the detailed evolution of individual events.

\subsubsection{Velocities as a function of phase}
\label{phase2}

The  \SiII\ 4130 \AA, \SiII\ 5972 \AA, \SiII\ 6355 \AA, and \Caii\ NIR triplet velocity (PVF and HVF) components are shown as a function of phase in Fig.~\ref{phase}. As expected, we find that the velocities of the spectral lines decrease with time over the phase range of $-18$ to +5 d. The slopes of the best-fit lines to the data in a maximum-light sample (phase range $-5$ to +5 d) are  $-77\pm28$, $+47\pm36$, $-60\pm25$, $+17\pm59$, $-310\pm75$ \kms\ d$^{-1}$ for \SiII\ 4130 \AA, \SiII\ 5972 \AA, \SiII\ 6355 \AA, \Caii\ NIR PVF, and \Caii\ NIR HVF, respectively. Previous studies such as \cite{ben05} and \cite{blo12} examined the detailed phase evolution of individual objects and found a diversity in the rate of the evolution of the line velocities with time, as well as sometimes a nonlinear evolution with time. Therefore, we do not correct our measured velocities to maximum light using these average best-fit lines but instead restrict the sample to phase ranges over which the evolution with time is expected to be small. 

In the maximum-light sample, the weighted mean \SiII\ velocities are $9941\pm84$, 10,511$\pm$ 108 and 11,477 $\pm$ 96 \kms\ for the \SiII\ 4130 \AA, \SiII\ 5972 \AA\ and \SiII\ 6355 \AA\ features, respectively. As expected, the weaker \SiII\ 4130 \AA\ and 5972 \AA\ lines have lower velocities than the stronger \SiII\ 6355 \AA. 

There is a significant velocity difference between the weighted mean photospheric velocity of the \Caii\ NIR triplet (10,085 $\pm$ 173 \kms) and the mean high-velocity \Caii\ NIR component (18,550 $\pm$ 151 \kms) near maximum light. In our fitting routine, the greatest forced difference between the velocity of the \Caii\ PVF component and velocity of the \Caii\ NIR `high-velocity' component is $\sim 3000$ \kms\ (and in many cases is significantly less). However, our measured average difference between the \Caii\ NIR components of $\sim 8500$ \kms\ is significantly larger than this value.  Further discussion of HVFs will be presented in Section \ref{hvf}.

\subsubsection{pEW as a function of phase}
\label{pew_phase}
The pEW of the main \SiII\ (4130 \AA, 5972 \AA, 6355 \AA) and the \Caii\ NIR lines in the SN Ia spectra are also investigated, and are shown as a function of phase in Fig.~\ref{phase}. There is a much larger scatter, extending to much higher values, in the \SiII\ 6355 pEW compared to the pEW of the \SiII\ 4130 and \SiII\ 5972 \AA\ lines. The standard deviations of the means are 7 \AA, 12 \AA\ and 37 \AA\ in the phase range $-5$ to +5 d for \SiII\ 4130 \AA, \SiII\ 5972 \AA, \SiII\ 6355 \AA, respectively. In the maximum-light sample, the pEW of the three \SiII\ lines stay relatively constant with phase. 

The pEW of the two \Caii\ IR triplet components (PVF and HVF) evolve differently with time. As shown in Fig.~\ref{phase}, over the phase range of $-18$ to +5 d, the HVF strength decreases rapidly while the PVF strength tends to increase slightly. This is not surprising:  the HVF form in lower-density regions farther out in the ejecta and thus, are likely to become optically thin more quickly than the PVF.

\begin{figure*}
\includegraphics[width=16cm]{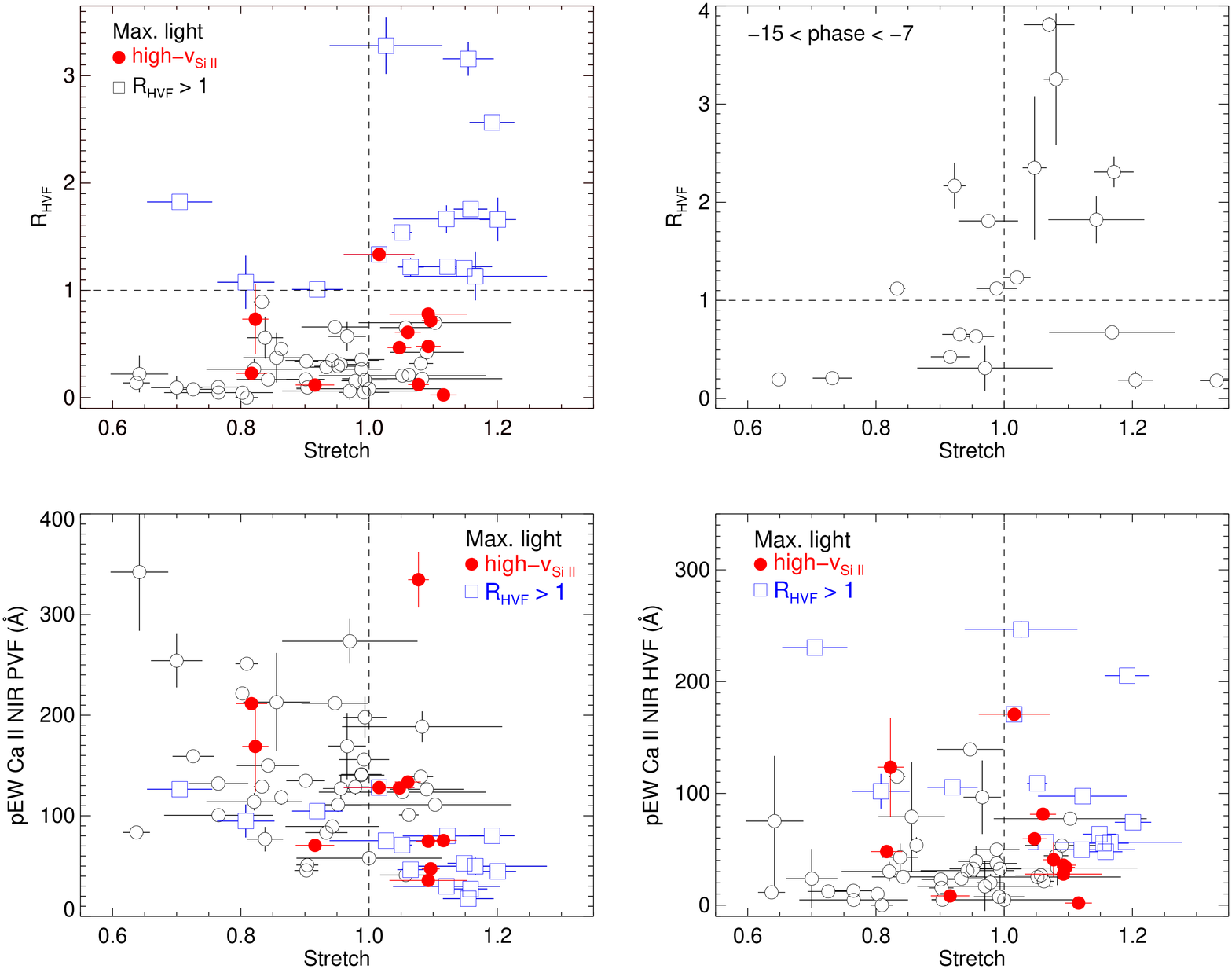}
\caption{\textit{Top-left panel:} $R_{\rm HVF}$ (ratio of the HVF to PVF pEW of the \Caii\ NIR triplet) as a function of the light-curve width, stretch in the phase range $-$5 to +5 d (open, black circles). Two dashed lines are marked at values of $s=1$ and $R_{\rm HVF}=1$. SNe Ia falling in the `high-\vsiii' \SiII\ 6355 \AA\ ($v>$\,12,000 \kms) subgroup are shown as red, filled circles. The SNe Ia in the maximum-light sample that have $R_{\rm HVF}>1$ are shown as open, blue squares. SNe Ia in the $R_{\rm HVF}>1$ subgroup have higher stretches, on average, than the rest of the sample, while the `high-\vsiii' \SiII\ 6355 \AA\ SNe Ia rarely have $R_{\rm HVF}>1$.   \textit{Top-right panel:} $R_{\rm HVF}$ as a function of the light-curve width, stretch in the phase range $-15$ to $-7$ d (open, black circles). \textit{Bottom-left panel:} pEW of the \Caii\ PVF against stretch for the $-5$ to +5 d sample. The SNe Ia in the maximum-light sample that have $R_{\rm HVF}>1$ are shown as blue, open squares. SNe Ia falling in the `high-\vsiii' \SiII\ 6355 \AA\ subgroup are shown as red, filled circles. A dashed line marks a value of $s=1$. \textit{Bottom-right panel:} pEW of the \Caii\ HVF as a function of stretch. The SNe Ia in the maximum-light sample that have $R_{\rm HVF}>1$ are shown as blue, open squares, while SNe Ia falling in the `high-\vsiii' \SiII\ 6355 \AA\ subgroup are shown as red, filled circles. A dashed line marks a value of $s=1$. }
\label{hvf_stretch}
\end{figure*}

\subsection{`Normal' and `High' velocities and light-curve width}
\label{normal_high}

Previous studies have suggested that SNe Ia can be split into two distinct populations based on their \SiII\ 6355 \AA\ velocities \citep{2009ApJ...699L.139W,2013Sci...340..170W}, `high-velocity' (high-\vsiii,  $v>$ 11,800--12,000 \kms) and `normal-velocity' (normal-\vsiii, $v<$11,800--12,000 \kms). To investigate the relationship between \SiII\ 6355 \AA\ velocity, the velocity of the \Caii\ NIR triplet components and light-curve width, we focus on a maximum-light sample of SNe Ia ($-5$ to +5 d with respect to \textit{B}-band maximum). We do not correct the velocities to maximum light for the reasons described in Section \ref{phase2}. If more than one spectrum for a SN Ia is present in this phase range, we choose the spectrum closest to maximum light.  Two SNe Ia in our sample appear as outliers owing to their unusually large light-curve widths of $s>1.4$ (PTF10glo and PTF09dhx) and are excluded from further analysis (their positions in \SiII\ 6355 \AA\ velocity vs. stretch space can be seen in the left panel of Fig.~\ref{col_str}). The spectrum of PTF10glo appears similar to normal SNe Ia, while the spectra of PTF09dhx appear similar to `super-Chandrasekhar' events \citep[e.g.,][]{2006Natur.443..308H,2007ApJ...669L..17H,2010ApJ...713.1073S,2011MNRAS.410..585S}.

The left plot of Fig.~\ref{col_str} shows the velocity of the \SiII\ 6355 \AA\ line as a function of stretch. We measure the statistical significance of potential trends between the \SiII\ 6355 \AA\ velocity and light-curve width using \textsc{LINMIX}, a Bayesian approach for linear regression with uncertainties in both variables \citep{2007ApJ...665.1489K}. As in previous studies, we find no correlation between the \SiII\ 6355 \AA\ velocity and light-curve width for this maximum-light sample, nor when we limit the sample to that used by cosmological studies \citep[$0.7<$ stretch $<1.3$;][]{2011ApJS..192....1C}. We also consider the \SiII\ 6355 \AA\ high-velocity and normal-velocity subsamples separately and look for correlations between light-curve width and \SiII\ 6355 \AA\ velocity within the subsamples, but again find no statistically significant trends. In Fig.~\ref{col_str} (middle panel), we show our measured \Caii\ NIR PVF and HVF velocities as a function of stretch.  Using \textsc{LINMIX}, we find no correlation between the \Caii\ NIR PVF velocity and light-curve width, and only a weak relation between the \Caii\ NIR velocity and light-curve width (2.8-$\sigma$).

 \begin{figure*}
\includegraphics[width=17.5cm]{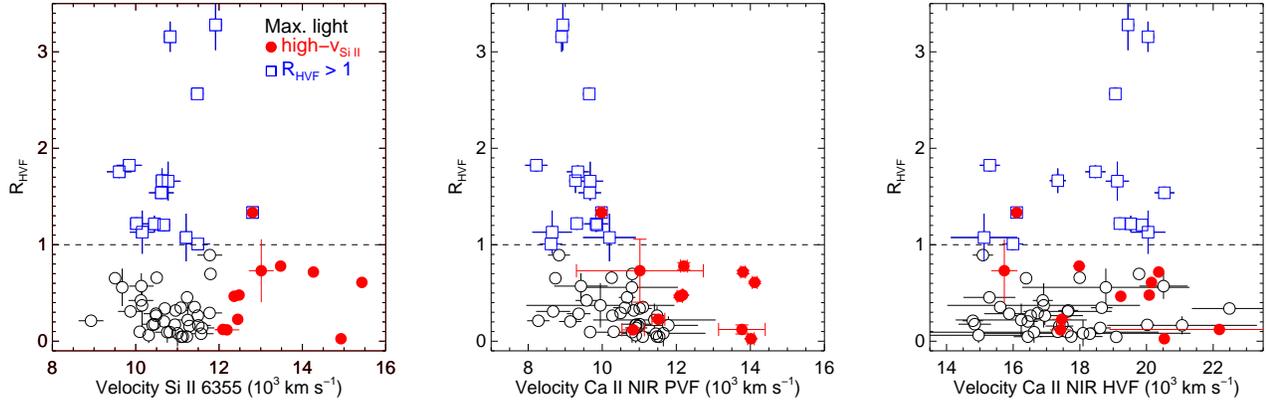}
\caption{$R_{\rm HVF}$ ratio against \SiII\ 6355 \AA\ velocity (left panel) in the phase range $-5$ to +5 d, as a function of \Caii\ NIR PVF velocity (middle panel) and as a function of \Caii\ NIR HVF velocity (right panel). A dashed horizontal line marks a value of $R_{\rm HVF}=1$. SNe Ia with $R_{\rm HVF}>1$ have lower \SiII\ 6355 \AA\ and  \Caii\ NIR PVF velocities compared to the $R_{\rm HVF}<1$ subset. SNe Ia falling in the `high-\vsiii' \SiII\ 6355 \AA\ group are shown as red, filled circles and SNe Ia that have $R_{\rm HVF}>1$ are shown as filled, blue squares.  }
\label{hvf_vel}
\end{figure*}

The \Caii\ NIR PVF and HVF velocities against the \SiII\ 6355 \AA\ velocity in the maximum-light sample are shown in the right panel of Fig.~\ref{col_str}.  A clear trend is seen between the \Caii\ NIR PVF and the \SiII\ 6355 \AA\ velocity, with a 8.8$\sigma$ significance. The corresponding Pearson correlation coefficient is 0.73. No correlation between the \Caii\ HVF velocity and the \SiII\ 6355 velocity is seen.  If we remove SNe Ia with `high-\vsiii,' then the correlation between the \Caii\ PVF velocity and the \SiII\ 6355 \AA\ velocity drops to 3.4-$\sigma$.  This suggests that the correlation between \Caii\ PVF and \SiII\ 6355 \AA\ velocity is at least partly driven by the `high-\vsiii' SNe Ia having higher \Caii\ PVF velocities.

In our fitting routine (as discussed in Section \ref{fitting}), we constrain the velocity of the \Caii\ NIR PVF to be within 25 per cent of the \SiII\ 6355 \AA\ velocity. This is a much looser constraint than that of \citet[][]{2014MNRAS.437..338C} of within 10 per cent of the \SiII\ 6355 \AA\ velocity. To test the effect of this on the relation between the \Caii\ PVF and \SiII\ 6355 \AA\ velocities, we test the less stringent constraints of the \Caii\ NIR PVF velocity being within 40 per cent and 60 per cent of the \SiII\ 6355 \AA\ velocity. We find no significant difference in the correlation values for these constraints. Therefore, the tight correlation between \SiII\ 6355 \AA\ velocity and the \Caii\ NIR PVF velocity cannot be explained by this constraint alone.  We discuss the possible origin of the observed variations in line velocities in Section \ref{siii_div}.

\subsection{Strength of high-velocity \Caii\ features}
\label{hvf}
High-velocity features of the \SiII\ 6355 \AA, \Caii\ H\&K and \Caii\ NIR triplet lines have been previously identified in many SN Ia spectra at early times and their origin is still unclear \cite[e.g.,][]{2013ApJ...770...29C,2004ApJ...607..391G,2005ApJ...623L..37M}. These features typically have velocities that are a few thousand \kms\ higher than the photospheric component (as seen for our sample in Fig.~\ref{phase} where we identify an average difference of $\sim 8500$ \kms) and can appear as multiple-peaked or blended absorption profiles. 

Previous studies have been made of the \Caii\ H\&K region because data were more readily available for it than the \Caii\ NIR triplet region. However, the \Caii\ NIR region is less affected by other lines and so is more suitable for studying the presence of high-velocity features. \cite{2014MNRAS.437..338C} parameterised the ratio of the HVF to the PVF pEW of the \Caii\ NIR triplet as $R_{\rm HVF}$. To investigate the presence and strength of HVF in the PTF SN Ia sample, we measure the pEW of the \Caii\ PVF and HVF in spectra in our sample that cover the \Caii\ NIR triplet and are in the phase range of $-$18 to +5 d with respect to \textit{B}-band maximum.  

 \begin{figure*}
\includegraphics[width=14.5cm]{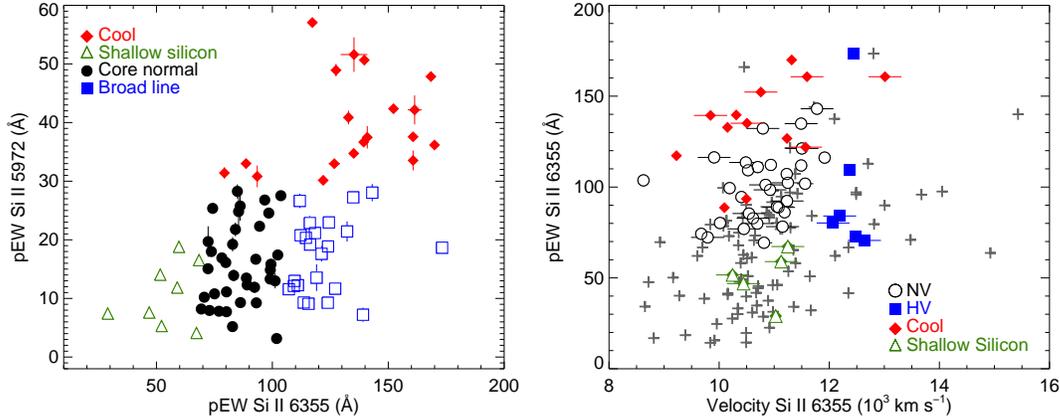}
\caption{\textit{Left:} pEW of the \SiII\ 5972 \AA\ against the pEW of the \SiII\ 6355 \AA\ absorption for SNe Ia with spectra in the phase range from $-5$ to +5 days with respect to maximum light. The sample is split into four subclasses: `cool' (red filled diamonds), `broad line' (blue open squares), `shallow silicon' (green open triangles) and `core-normal' (black filled circles) as defined by \protect \cite{2006PASP..118..560B}. The subclasses are colour-coded the same as fig.~8 of  \protect \cite{blo12}. If more than one measurement was available in this phase range, the one closest to maximum light was used. \textit{Right:} pEW of the \SiII\ 6355 \AA\ feature as a function of the velocity of the \SiII\ 6355 \AA\ feature for SN Ia spectra within 3 days of maximum light, similar to fig.~2 of \protect \cite{2009ApJ...699L.139W}. If more than one measurement was available, the one closest to maximum light was used. The sample has been split into those with `normal-velocity' (normal-\vsiii; black open circles), `high-velocity' (high-\vsiii; blue filled squared), `cool' (red filled diamonds) and `shallow silicon' (green open triangles) as defined by  \protect  \cite{2009ApJ...699L.139W}. Grey `plus' symbols represent the SNe for which a subclass could not be defined.}
\label{class_plot}
\end{figure*}

\subsubsection{\Caii\ HVF strength and light-curve width}
\label{hvf_lightcurve}
In Fig.~\ref{hvf_stretch}, we show the \Caii\ $R_{\rm HVF}$ against the light-curve width (stretch) in two different phase bins, $-$5 to +5 d (maximum light) and $-$15 to $-$7 d (early). We confirm the maximum-light relationship identified by  \cite{2014MNRAS.437..338C}, as well as show that this relation is present in our early time sample.  We show that SNe Ia with broader light curves (higher stretch) have higher values of $R_{\rm HVF}$, on average. To estimate the significance of this relationship, we split the maximum-light sample in two at $R_{\rm HVF}=1$ (where the strength of both components is equal), and find that the weighted mean stretch of the low $R_{\rm HVF}$ is $0.93\pm0.02$, while the high $R_{\rm HVF}$ weighted mean stretch is $1.08\pm0.03$. This corresponds to a difference in their weighted mean stretches of 3.3$\sigma$. 

To better understand the origin of the trend between $R_{\rm HVF}$ and stretch, we have studied the pEW of the individual photospheric and high-velocity components as a function of stretch (bottom panels of Fig.~\ref{hvf_stretch}). We find that the weighted means of the PVF pEW for the $R_{\rm HVF}>1$ and $R_{\rm HVF}<1$ subsamples are different at the 3.6$\sigma$ level, with mean values of $59\pm9$ \AA\ and $121\pm8$ \AA, respectively. For the HVF, we find the weighted mean pEW of the $R_{\rm HVF}>1$ and $R_{\rm HVF}<1$ subsamples to be different at the 5.4$\sigma$ level, with mean values of $133\pm19$ \AA\ and $16\pm3$ \AA.  This suggests that there are two effects driving the $R_{\rm HVF}$ versus stretch relation -- there is both a lower contribution from the \Caii\ NIR photospheric velocity components in higher-stretch SNe Ia, as well as stronger \Caii\ HVF in more luminous SNe Ia. We will discuss possible reasons for this in Section \ref{origin}.

\subsubsection{\Caii\ HVF strength and spectral velocities}
\label{hvf_specvel}

We also investigate the \Caii\ NIR $R_{\rm HVF}$ as a function of the velocities of the spectral features.   In Fig.~\ref{hvf_vel}, we show the $R_{\rm HVF}$ as a function of the \SiII\ 6355 \AA, \Caii\ NIR PVF and \Caii\ NIR HVF velocities in the maximum-light sample. While only one SN Ia in our high-$R_{\rm HVF}$ sample has a `high-\vsiii' \SiII\ 6355 \AA\ value, we do not find a statistically significant difference in the weighted mean \SiII\ velocities for the high-$R_{\rm HVF}$ and low-$R_{\rm HVF}$ subsamples.

However, we have also investigated the relationship between the \Caii\ NIR $R_{\rm HVF}$ and the velocity of the \Caii\ NIR components. We find that SNe Ia with high $R_{\rm HVF}$ values have higher \Caii\ photospheric velocities compared to the low $R_{\rm HVF}$ sample (different at the 12.6$\sigma$ level) -- the weighted means of the \Caii\ NIR PVF velocities for the high-$R_{\rm HVF}$ and low-$R_{\rm HVF}$ subsamples are $8947\pm23$ \kms\ and 11,265 $\pm$ 161 \kms, respectively. For the high-velocity \Caii\ NIR component, we find no significant difference in the weighted mean velocities for the high-$R_{\rm HVF}$ and low-$R_{\rm HVF}$ subsamples.

\subsection{Spectroscopic subclasses}
\label{subclasses}

Different classification systems for splitting SNe Ia into spectral subclasses have previously been defined \citep{2006PASP..118..560B,2009ApJ...699L.139W}. In Fig.~\ref{class_plot} (left panel), the pEW \SiII\ 5972 \AA\ against pEW \SiII\ 6355 \AA\ of \cite{2006PASP..118..560B} in the phase range $-5$ to +5 d is shown. If more than one measurement was available in this phase range, the measurement closest to maximum light was used. The pEW \SiII\ 5972 \AA\ against pEW \SiII\ 6355 \AA\ space is divided into four subclasses: `cool' (high pEW \SiII\ 5972 \AA\ values, SN 1991bg-like objects), `shallow silicon' (both shallow \SiII\ 5972 \AA\ and \SiII\ 6355 \AA, SN 1991T-like objects), `core normal' (middle range of pEW values, `normal' SNe Ia), and `broad line' (SNe with normal pEW \SiII\ 5972 but higher \SiII\ 6355 \AA\ widths). 
`Cool' SNe Ia also display \TIii\ absorption at $\sim 4200$ \AA, which is a signature of lower temperatures and less luminous SNe Ia. 
Of the 86 SNe Ia for which both the pEW \SiII\ 5972 \AA\ and pEW \SiII\ 6355 \AA\ could be measured in the phase range $-5$ to +5 d , we find 36 SNe Ia ($\sim$42 per cent) that are classified as core-normal, 23 SNe Ia ($\sim 27$ per cent) as `broad line,' 8 SNe Ia ($\sim 9$ per cent) as `shallow silicon' and 19 SNe Ia ($\sim 22$ per cent) as `cool.' The number of SNe Ia falling in the `core-normal,' `broad line' and `cool' subclasses are within $\sim 25$ per cent of the values found by \cite{blo12} for the CfA SN Ia spectral sample. A larger difference is seen for the `shallow silicon' subclass, where \cite{blo12} found 14 per cent of the SNe fell in it. However, given the relatively small number of objects in this subclass, the difference is not significant.

In the right panel of Fig.~\ref{class_plot}, the pEW of the \SiII\ 6355 \AA\ line against its velocity in the phase range $-$3 to +3 d, as presented by \cite{2009ApJ...699L.139W}, is shown. Unlike \cite{2009ApJ...699L.139W}, we do not split the sample into subclasses based on a qualitative visual inspection of the spectra (SN 1991T-like, SN 1991bg-like, `normal'). Instead, we use their position in the pEW \SiII\ 5972 \AA\ against pEW \SiII\ 6355 \AA\ space (left panel of Fig.~\ref{class_plot}) to provide a quantitative measure of their peculiarities. We consider the `core-normal' and `broad-line' subgroups to be `Branch-normal' SNe Ia as generally defined in the literature \citep[see][]{2006PASP..118..560B}, while the `cool' group includes SN 1991bg-like objects and the `shallow silicon' group includes SN 1991T-like objects, but there is not a one-to-one overlap. After excluding SNe Ia falling in the `shallow silicon' and `cool' subgroups of \cite{2006PASP..118..560B}, we follow \cite{2013Sci...340..170W} and define any of the remaining SNe Ia as  `Branch-normal' SNe Ia, and split them into `high-\vsiii' ($>$\,12000 \kms) and `normal-\vsiii' ($<$\,12000 \kms). A \cite{2006PASP..118..560B} subclassification could not be made for all the SNe Ia mainly because the S/N was too low in the \SiII\ 5972 \AA\ region or the feature was contaminated by host-galaxy lines. 

\subsection{Spectral luminosity indicators}
\label{sec:lum_in}

\begin{figure*}
\includegraphics[width=14.5cm]{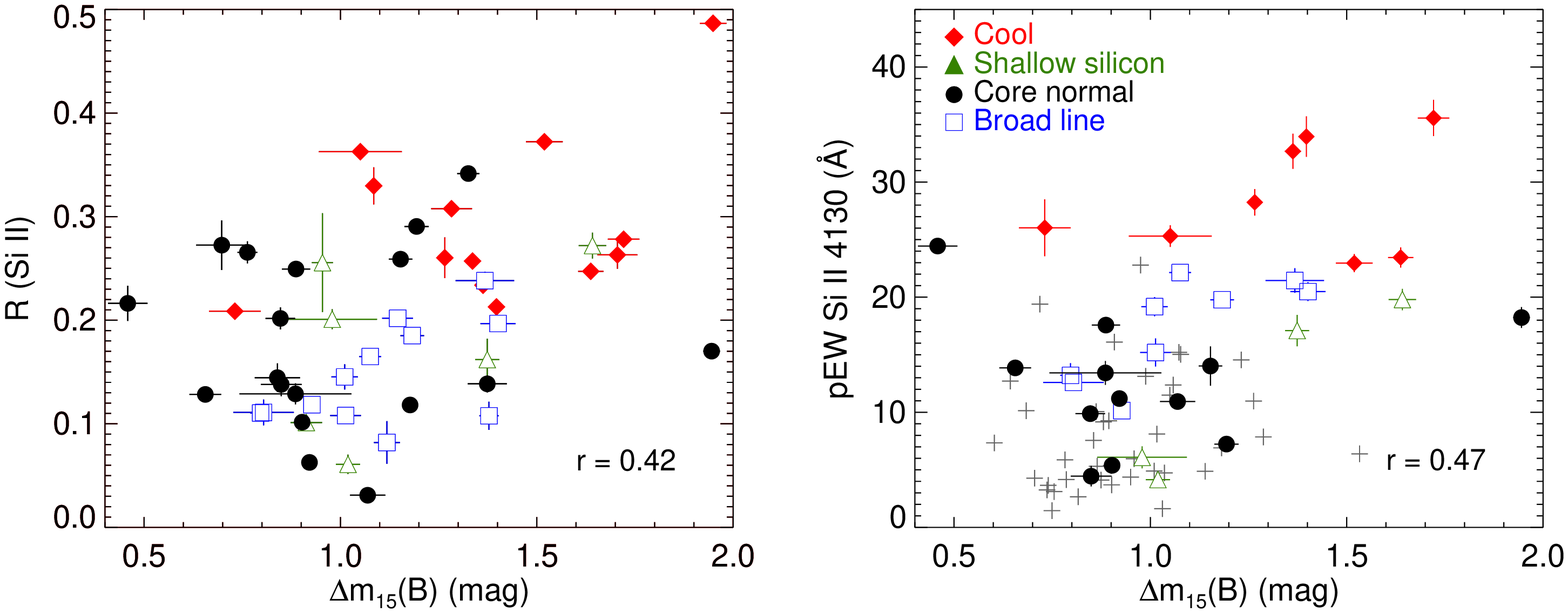}
\caption{Luminosity indicators for SNe Ia: R(\SiII), ratio of the pEW of \SiII\ 5972 \AA\ to pEW of \SiII\ 6355 \AA\ line against stretch (left panel) and pEW of \SiII\ 4130 \AA\ line against stretch (right panel) in the phase range -2.5 to +2.5 d with respect to maximum light \protect \cite[following][]{2011A&A...526A..81B}. The spectroscopic subclasses of  \protect \cite{2006PASP..118..560B} are marked for each SN as in Fig.~\ref{class_plot}, grey `plus' symbols represent the SNe for which a subclass could not be defined. The Pearson coefficients for each plot for the whole samples are marked. }
\label{lum_in}
\end{figure*}

Previous studies have investigated the relation between SN Ia spectral features and light-curve parameters with the aim of improving the use of SNe Ia as distance indicators. 
In Fig.~\ref{lum_in}, we have plotted measurements of two well-studied spectral luminosity indicators against light-curve width (stretch): the ratio of the pEW of \SiII\ 5972 \AA\ to pEW of \SiII\ 6355 \AA\ line as defined by \cite{2008MNRAS.389.1087H}, which is an updated version of the R(\SiII) of \cite{1995ApJ...455L.147N}, and the pEW of the \SiII\ 4130 \AA\ line.  Following \cite{2011A&A...526A..81B}, we restrict our sample to spectra in the phase range of $-2.5$ to +2.5 d with respect to maximum light. We split the sample into subclasses (`cool,' `shallow silicon,' `core normal,' `broad line') as described in Section \ref{subclasses}. For a SN to be placed in one of these spectral luminosity subclasses, it is necessary that the pEW of the \SiII\ 5972 \AA\ and the \SiII\ 6355 \AA\ lines can be measured. 

For comparison with previous work, we convert our SiFTO `stretch' values to $\Delta m_{15}(B)$ \citep{1993ApJ...413L.105P} using the relation given by \cite{2008ApJ...681..482C}. Firstly, we investigate the connection between R(\SiII) and $\Delta m_{15}(B)$, finding a Pearson coefficient of 0.42 (Spearman rank coefficient of 0.36). For pEW of the \SiII\ 4130 \AA\ line and $\Delta m_{15}(B)$, we find a Pearson coefficient of 0.47 (Spearman rank coefficient of 0.43) for the whole sample.  These correlation values are of lower significance than previously found - revised values of \citet{blo12}\footnote{Some objects were accidentally left out of the original luminosity indicator analysis (Blondin; private communication)} for the R(\SiII) and the pEW of the \SiII\ 4130 \AA\ line against $\Delta m_{15}(B)$ are 0.67 and 0.66, respectively. \cite{2012MNRAS.425.1889S} find values of the correlation coefficient (Spearman rank coefficient) of $-$0.42 and $-$0.87 for the R(\SiII) and the pEW of the \SiII\ 4130 \AA\ line against the SALT2 light-curve width parameter, `x$_1$' (respectively). 

We also investigated the correlation between the most promising luminosity indicator (visible to higher redshifts than R(\SiII)), the pEW of the \SiII\ 4130 \AA\ line and stretch for members of the individual subclasses. We find that when the sample is restricted to SNe Ia falling in the `broad line' subclass alone (just 9 objects), the Pearson coefficient is $0.78\pm0.16$, meaning a significant correlation between the \SiII\ 4130 \AA\ pEW of `broad line' SNe Ia and light-curve width. 

 \begin{figure}
\includegraphics[width=7.5cm]{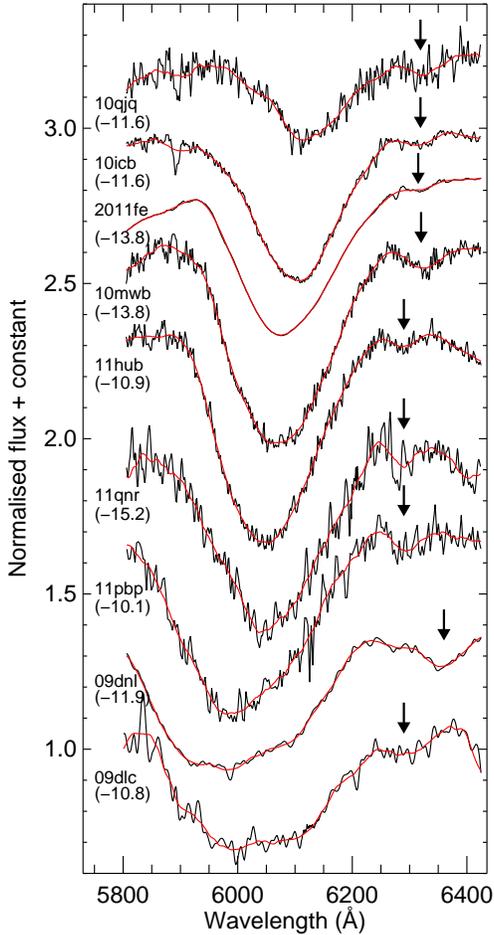}
\caption{Identification of a \Cii\ 6580 \AA\ feature in 9 SNe Ia having spectra before $-10$ d with respect to maximum light. SN 2011fe was previously shown to have \Cii\ 6580 \AA\ features in multiple early-time spectra \protect \citep{2012ApJ...752L..26P}. The black lines display the raw spectra, while the red lines show the spectra smoothed with a Savitzky-Golay filter \protect \citep{1964AnaCh..36.1627S}. The black arrows mark the position of maximum \Cii\ absorption.}
\label{Cdetect_plot}
\end{figure}

\subsection{Searching for carbon at early times}
\label{Carbon}

The presence of C in SN Ia spectra is an excellent tracer of unburnt material, and provides information on the way the star exploded. To investigate the presence of \Cii\ 6580 \AA\ at early times in our sample, we select SNe Ia with spectra at phases before $-10$ d with respect to maximum light, resulting in a sample of 24 SNe Ia with at least one spectrum in this phase range. Our sample includes 7 spectra of SN 2011fe \citep{2011Natur.480..344N,2012ApJ...752L..26P}. SN 2011fe was found to have clear signatures of \Cii\ detected in its early-time spectra lasting until nearly maximum light, with strongly decreasing strength with time \citep{2012ApJ...752L..26P}.  Two other SNe Ia in our sample also have 2 spectra each (PTF10zdk, PTF11hub).  Our sample also includes three spectra of SN 2010jn \cite[PTF10ygu;][]{2013MNRAS.429.2228H} where \Cii\ 6580 \AA\ was not detected. SN 2010jn was an unusual SN Ia with very broad \SiII\ features, and the highest \SiII\ velocities of our sample (Fig.~\ref{phase}). If the \Cii\ 6580 \AA\ velocity was also toward the higher end of \Cii\ velocities then it could be blended with the \SiII\ 6355 \AA\ features, resulting in a null detection \cite[for a discussion, see][]{2011ApJ...743...27T}.

We follow the nomenclature of \cite{2012ApJ...745...74F}, defining the detection of \Cii\ 6580 \AA\ as `absorption,' `flat,' or `no detection' based on a visual inspection of the spectra. A spectrum in the `absorption' class shows a clear absorption feature at the expected position of \Cii\ 6580 \AA\ ($\sim 6280$--6350 \AA, assuming typical velocities of $\sim$\,10,400--13,500 \kms), while the `flat' profile is one that appears to have a suppressed \SiII\ 6355 \AA\ emission resulting in a flat profile in the \Cii\ region.   A `no detection' label means straightforwardly that no flat or absorption profile is found. A `?' symbol after the initial classification means a low S/N in the \Cii\ region and an uncertain feature classification. 

In our sample, we find 9 SNe Ia (including SN 2011fe) with a definite `absorption' feature at the position of \Cii\ 6580 \AA\, with an additional two SNe Ia having an `absorption?' feature (Table \ref{tab:Cdetect}). Fig.~\ref{Cdetect_plot} shows the SNe Ia where \Cii\ 6458 \AA\ has been positively detected in their spectra.  Two SNe Ia are classified as `flat' profiles, two as `flat?,' six as `no detection,' and three SNe Ia fall in the `no detection?' subclass. These results suggest that a lower limit of $\sim 38$ per cent of SNe Ia show \Cii\ features (i.e., with an `absorption classification') if observed at earlier than $-10$ d. 

Similarly to the trend first identified by \cite{2011ApJ...743...27T}, we also find a correlation between light-curve width and the presence of \Cii, in the sense that SNe Ia with \Cii\ features have narrower light curves on average. The weighted mean stretch distribution for the `absorption' sample is $0.92\pm0.03$ and for the `no detection' sample is $1.11\pm0.03$, different at a significance $\sim$3.5$\sigma$.

We also investigate the connection between \Caii\ NIR $R_{\rm HVF}$ and the presence of \Cii\ 6580 \AA\ -- 12 of the SNe Ia with a spectrum before $-10$ d have a $R_{\rm HVF}$ measurement around maximum light. Of the six SNe Ia in the sample in the `absorption' group that  have a measured value of $R_{\rm HVF}$, five have $R_{\rm HVF}<1$ and one (PTF09dlc) has $R_{\rm HVF}>1$. The three SNe Ia in the sample in the `no detection' C group have $R_{\rm HVF}<1$. Therefore, we do not identify any trends between $R_{\rm HVF}$ and detection of \Cii\ 6580 \AA\ in this small sample.

\begin{table}
  \caption{\Cii\ 6580 \AA\ classification for 24 SNe Ia having at least one spectrum earlier than $-10$ d with respect to \textit{B}-band maximum.}
 \label{tab:Cdetect}
\begin{tabular}{@{}lccccccccccccccccccccccccccccc}
  \hline
  \hline
\Cii\ class &No. of SNe Ia & \% of total sample\\
\hline
\hline
Absorption&9&37.5 \\
Flat emission&2 &8.3\\
No detection&6&25.0\\
Absorption?&2&8.3\\
Flat emission?&2&8.3\\
No detection?&3&12.5\\
     \hline
\end{tabular}
 \begin{flushleft}
\end{flushleft}
\end{table}

\section{Discussion}
\label{discussion}

We have presented spectra of 264 SNe Ia having at least one spectrum before +5 d from the PTF SN Ia programme. These data were combined with light-curve width information to provide a new low-redshift SN Ia dataset (unbiased with respect to host-galaxy properties) from which the diversity of SNe Ia can be investigated and correlations between spectral properties, as well as with light-curve width, can be determined. An accompanying analysis of the PTF spectral sample combined with host-galaxy information is given by Pan et al. (submitted). 

\subsection{The diversity of \SiII\ velocities}
\label{siii_div}
The connection between spectral velocities and light-curve properties provides important information on the configuration of the explosion. Spectral feature velocities provide us with information on the location of line-forming regions. We find that mean velocities for the weaker \SiII\ 4130 and \SiII\ 5972 \AA\ features are lower than for the stronger \SiII\ 6355 \AA\ line, whose line-forming region extends to higher velocity.

\cite{2009ApJ...699L.139W} suggested that there are two distinct spectroscopic subclasses for SNe Ia: those with \SiII\ 6355 \AA\ velocities above 11,800 \kms\ at maximum light, and those with values below. \cite{2013Sci...340..170W} investigated the galaxy environments of these SNe Ia with high \SiII\ 6355 \AA\ velocities and fo und that they occur preferentially in more massive host galaxies and in the central regions of their hosts, compared to SNe Ia with lower \SiII\ 6355 \AA\ velocities that tend to occur at larger radii. About 33 per cent of \cite{2013Sci...340..170W} sample fall in the high-velocity subclass ($>$\,12,000 \kms). For our sample, we find significantly fewer SNe Ia in this high-velocity subgroup ($\sim 16$ per cent). A study focussing on the connection between host-galaxy properties and spectral features in the PTF sample (Pan et al. submitted) find that this is at least partially caused by selection biases in the LOSS SN sample that results in more massive host galaxies compared to our SN sample  (although this does not  completely explain the different number of high-\vsiii\ events found). 

We have identified a strong relation between the \SiII\ velocity and the velocity of the \Caii\ NIR PVF. SNe Ia with higher \SiII\ velocities have higher \Caii\ NIR PVF velocities (8.8-$\sigma$). This result is in agreement with both one-dimensional deflagration and delayed-detonation models \cite[e.g.,][]{1999ApJS..125..439I,2013MNRAS.429.2127B,2014MNRAS.439.1959M} showing that the distribution of intermediate-mass elements such as Si and Ca in the ejecta is roughly similar.

High-velocity features in the \SiII\ 6355 \AA\ feature were first suggested to be present at early times in SN 1990N \citep{2001MNRAS.321..341M}, and since then have been identified in a number of SNe Ia \citep{2011Natur.480..344N,2012ApJ...752L..26P,2013ApJ...770...29C}. A contribution from a high-velocity component in the \SiII\ 6355 \AA\ line may be present but blended with the photospheric component, as in SN 1990N, or isolated, as in the case of SN 2011fr. One way of testing for the presence of high-velocity features in the \SiII\ lines is to compare their velocities to the \Caii\ NIR PVF velocity: both are intermediate-mass elements and should have similar mass distributions in the ejecta. We find in an early-time sample ($-$15 to $-$7 d), for which both \Caii\ NIR PVF and \SiII\ 6355 \AA\ velocity measurements are available, the velocity of the \SiII\ 6355 \AA\ line is higher than the \Caii\ NIR PVF velocity at the 2.6-$\sigma$ level. This suggests that there may be a weak contribution from high-velocity components to the strong \SiII\ 6355 \AA\ feature at early times.

\subsection{The origin of high-velocity \Caii\ features}
\label{origin}
 One of the main foci of this work is to study the presence of high-velocity features in the \Caii\ NIR triplet and measure their contribution to the overall profile.  In many cases, the high-velocity features are blended with their corresponding photospheric components. However, they can be disentangled using multi-Gaussian spectral fitting techniques. The physical origin for high-velocity features is unclear -- the two main interpretations for these high-velocity features are either an abundance or density enhancement at large radii. An abundance enhancement of Ca could cause these high-velocity features but would need very large enhancements to explain the strong high-velocities features seen in some SNe Ia \citep{2005ApJ...623L..37M}. A density enhancement could arise from either circumstellar material or could be intrinsic to the SN, coming from material in the outer layers \citep{2005ApJ...623L..37M,2008ApJ...677..448T}. Mixing of H from CSM can increase the rate of recombination resulting in stronger \Caii\ features at high velocity \citep{2005ApJ...623L..37M,2008ApJ...677..448T}. We have found that the mean \Caii\ NIR HVF velocity decreases significantly faster than the mean PVF velocity from $-$18 d to +5 d. The faster evolution of the HVF is most easily explained by HVF becoming optically thin quicker than the PVF since it is formed in a lower density region farther out in the ejecta, as discussed in \cite{2005ApJ...623L..37M}.
 
 The HVF velocities are found to be significantly higher than the PVF velocities for our sample, with a mean velocity difference between the two features of $\sim 8500$ \kms. This is much higher than the greatest difference enforced by the fitting routine of $\sim 3000$ \kms. This large mean velocity difference suggests that the material that gives rise to the \Caii\ HVF should be located at the edge of the ejecta. It may be part of the ejecta from the explosion, or could be due to a small mass of CSM dragged by the exploding material. We do not identify a trend between \SiII\ 6355 \AA\ velocity and the velocity of the \Caii\ NIR HVF. If these velocities were connected, then this could suggest the  \Caii\ HVF must be intrinsic to the SN. However, since they are unconnected, the possibility that the origin of the \Caii\ HVF is external to the SN (from CSM) cannot be ruled out.
 
\subsubsection{How common are high-velocity \Caii\ features?}

We wish to estimate how common are high-velocity \Caii\ features in SNe Ia. As expected, a contribution from a \Caii\ photospheric component is necessary to fit the \Caii\ NIR feature for all the SNe Ia in our sample. We estimate the percentage of SNe Ia that require a contribution to the \Caii\ NIR triplet from a high-velocity component by calculating the number of SNe Ia with a HVF pEW that is greater than the minimum PVF pEW in the same phase range. This is equivalent to assuming that if the pEW of the component is lower than the lowest PVF pEW, it is essentially negligible, while conservatively allowing for small nonzero HVF pEW in the fitting routine.

At early phases (before $-5$ d), we find that $\sim 95$ per cent of SNe Ia (18 out of 19 SNe Ia) require a contribution from a high-velocity \Caii\ NIR feature. This drops to $\sim 80$ per cent in the maximum-light sample ($-$5 to +5 d), showing that HVF are more prevalent at early times compared to maximum light.  This demonstrates that nearly all SNe Ia have high-velocity \Caii\ NIR features at early times, confirming the suggestion of \cite{2005ApJ...623L..37M}.

\subsubsection{Relative strength of high-velocity \Caii\ features}

 The $R_{\rm HVF}$ of the \Caii\ NIR triplet was defined by \cite{2014MNRAS.437..338C} to quantify the strength of HVF relative to the PVF in the \Caii\ NIR triplet. They found, in a maximum-light sample, that SNe Ia with higher $R_{\rm HVF}$ values have broader light curves. We confirm this trend in our sample (Fig.~\ref{hvf_stretch}), and also show the trend is present at earlier phases.
 At early times ($-$15 to $-$7 d), we find that $\sim 50$ per cent of SNe Ia have $R_{\rm HVF}$ values of greater than one (i.e., where the \Caii\ HVF dominates over the \Caii\ PVF), decreasing to $\sim 24$ per cent of SNe Ia in our maximum-light sample. There does not appear to be a one-to-one correlation between the $R_{\rm HVF}$ and the light-curve width, but instead an additional source of objects that is contributing, with $R_{\rm HVF}$ greater than one and light-curve widths of $s>1$ (as shown in Fig.~\ref{hvf_stretch}).  When the sample was split in two bins (high $R_{\rm HVF}$ and low $R_{\rm HVF}$), we found that there was a 3.3$\sigma$ difference between the weighted mean stretches.
 
We have also investigated the origin of this connection by studying the individual \Caii\ NIR HVF and PVF pEW as a function of light-curve width. It was found that dominant driver of the observed relation between $R_{\rm HVF}$ and stretch is that more luminous SNe Ia have higher pEW of their \Caii\ NIR HVF. This provides a link between more luminous SNe Ia and the presence of more high-velocity material, which could be caused by CSM or be intrinsic to the SN ejecta.

The more luminous SNe Ia are also found to have, on average, weaker PVF pEW. A possible reason for this is that in more luminous SNe Ia there is more burning of intermediate-mass elements in more luminous (broader light curve) SNe Ia \citep{2007Sci...315..825M}, resulting in less \Caii\ at the location of the photosphere, and hence a weaker PVF pEW. However, in this scenario, the photosphere may be expected to be located farther out in more luminous SNe Ia (higher \Caii\ PVF velocity), which was not seen for our sample in Section \ref{normal_high}. Another possibility is that the \Caii\ photospheric component feature is weaker in more luminous SNe Ia because of a higher Ca ionisation fraction in these events.  For the higher HVF pEW in more luminous SNe Ia, this could possibly be explained by an additional contribution at high-velocity to the SN Ia feature, which could be caused by CSM or be intrinsic to the SN ejecta.
 
 As shown by  \cite{2014MNRAS.437..338C}, SNe Ia appear to have either high $R_{\rm HVF}$ values \textit{or} high \SiII\ 6355 \AA\ velocities but not both (Fig.~\ref{hvf_vel}) -- we find only one SN Ia overlapping in these subsamples. However, the weighted mean \SiII\ velocities between the high-$R_{\rm HVF}$ and low-$R_{\rm HVF}$ subsamples is not significant. We also identified a connection between \SiII\ 6355 \AA\ velocity and \Caii\ NIR PVF velocity - SNe Ia with high \SiII\ 6355 \AA\ velocities have, on average, high \Caii\ NIR PVF velocities but rarely a high $R_{\rm HVF}$ value.  Pan et al. (submitted) studies the relation between the \Caii\ $R_{\rm HVF}$ and host-galaxy properties, finding that SNe Ia with high values of $R_{\rm HVF}$ are found preferentially in lower stellar mass hosts, while SNe Ia with high-\vsiii values are preferentially found in higher stellar mass populations.
 
 This could suggest that these two groups, SNe Ia with high $R_{\rm HVF}$ and SNe Ia with high \SiII\ velocities, are distinct and not generally overlapping in their spectral properties. The SNe Ia with high $R_{\rm HVF}$  values have lower \Caii\ PVF velocities, lower \SiII\ velocities, broader light curves (stretch $>1$) and occur in lower stellar mass hosts. Those in the high-velocity \SiII\ 6355 \AA\ subgroup have higher \Caii\ PVF velocities, lower $R_{\rm HVF}$ values, a range of light-curve widths ($0.76<$ stretch $<1.14$) and tend to occur in higher stellar mass host galaxies. The effect of observing an asymmetric explosion from different viewing angles can not explain these observations, since then there would be no connection between the presence (or absence) of \Caii\ $R_{\rm HVF}$ and high-\vsiii values and host-galaxy properties.

 \cite{2013MNRAS.436..222M} showed that SNe Ia showing blueshifted narrow \Nai\ D features, thought to be indicative of circumstellar material \citep[e.g.,][]{2007Sci...317..924P,2011Sci...333..856S}, also have broader light curves and occur more frequently in late-type galaxies. Some SNe Ia with blueshifted \Nai\ D features in the sample also fall in the high-velocity \SiII\ 6355 \AA\ subgroup, but not all \citep{2012ApJ...752..101F,2013MNRAS.436..222M}. However, the high-velocity \SiII\ 6355 \AA\ subgroup have a preference for occurring in the centre of galaxies and in more massive host galaxies \citep{2013Sci...340..170W} at odds with the preference of SNe Ia with blueshifted \Nai\ D features to occur late-type galaxies (low stellar mass, high specific star formation rate). Therefore, further investigation between the presence of blueshifted \Nai\ D features and \Caii\ HVF is necessary.

\subsubsection{\Caii\ velocity and light-curve parameters}
A correlation between \Caii\ velocity and light-curve width (SNe Ia with broader light curves having higher \Caii\ velocities) was previously identified for the \Caii\ H\&K feature \citep{1994AJ....108.2233W,1995ApJ...447L..73F,2012MNRAS.426.2359M,2013MNRAS.435..273F}. The origin of the correlation was questioned because the \Caii\ H\&K region may be contaminated by weaker lines such as \SiII\ 3858 \AA. However, \cite{2014MNRAS.437..338C} showed that the strength of high-velocity components relative to their photospheric components in the cleaner \Caii\ NIR feature is stronger in SNe Ia with broader light curves, which could therefore, influence the measured \Caii\ velocities and produce the observed trends.

For our sample, we analysed the individual  \Caii\ NIR HVF and PVF velocities as a function of stretch and find a weak correlation (2.8-$\sigma$) between the \Caii\ NIR HVF and the light-curve width. This suggests that previously observed trends  in the \Caii\ H\&K region (measured using a single Gaussian fit) may be somewhat driven by a higher \Caii\ HVF velocity in more luminous SNe Ia, but is likely also driven by the higher relative strength of \Caii\ HVF in more luminous SNe Ia.
 
\subsection{The future of SN Ia spectral luminosity indicators}
Recent studies have shown correlations between spectral luminosity indicators and light-curve width in SNe Ia at varying degrees of significance.  No studies combining both photometric and spectroscopic data have dramatically increased the performance of purely photometric data. However, studies have uncovered spectral luminosity indicators that can be used to standardise SN Ia luminosities, with varying levels of significance  \citep[e.g.,][]{1995ApJ...455L.147N,2008A&A...477..717B,2008MNRAS.389.1087H,2009A&A...500L..17B,2011A&A...526A..81B,2011MNRAS.410.1262W,2011ApJ...734...42N,2011A&A...529L...4C,2012MNRAS.425.1889S,blo12}. The main cause of this lack of an improvement appears to be that the light-curve width and spectral luminosity indicators such as R(\SiII) and the \SiII\ 4130 \AA\ pEW appear to trace different properties of the SNe Ia, with SNe Ia with very different light-curve widths having similar \SiII\ 4130 \AA\ pEW values.

We find, on average, lower significance trends for both the R(\SiII) and the \SiII\ 4130 \AA\ pEW against light curve stretch than previous studies for a number of possible reasons. The most likely reason is the increased number of SNe Ia with unusual properties that are present in the PTF SN Ia sample. However, as noted by \cite{blo12}, by selecting SNe Ia in just one of the subclasses of \cite{2006PASP..118..560B}, the significance of the trends can be increased. When we select only SNe Ia falling in the `broad line' subclass of \cite{2006PASP..118..560B}, the significance of the correlation is improved and we find for this subclass a Pearson correlation of $0.78\pm0.16$ for the \SiII\ 4130 \AA\ pEW against light-curve stretch. Therefore, with careful selection of the objects is it is possible to obtain a reasonable correlation between \SiII\ 4130 \AA\ pEW and light-curve width for our sample, but similarly to previous studies, not significantly improve upon the use of photometric data. 

Given the increased difficulty of obtaining spectra compared to obtaining photometry and the observed scatter in the correlation, the future of spectral luminosity indicators for SNe Ia in their current form is likely to be limited. However, ongoing spectroscopic surveys such as Nearby Supernova Factory \citep{2002SPIE.4836...61A,2004SPIE.5249..146L} that are obtaining high-quality multi-epoch spectra of a large sample of SNe Ia may find more sophisticated ways to connect spectral features to luminosity instead of simple pEW measurements and ratios.

\subsection{Ubiquity of \Cii\ and connection to progenitor models}

Identifying the presence of \Cii\ in SN Ia spectra is a key diagnostic of unburnt material in the SN ejecta, which provides important links to SN Ia explosion models, with different models predicting different amounts of \Cii\ left over in the outer layers.  Systematic searches for C in the early-time spectra of SNe Ia, when the outer layers are still visible, have been performed \citep{2007ApJ...654L..53T,2011ApJ...743...27T,2011ApJ...732...30P,2012ApJ...745...74F,2012MNRAS.425.1917S}. The most recent results of \cite{2012ApJ...745...74F} and \cite{2012MNRAS.425.1917S} have determined that $>$30--40 per cent of SNe Ia with spectra around $-10$ d have a clear detection of \Cii\ 6580 \AA\ in their spectra, and is suspected to be even higher when spectra earlier than $-$10 d are used \citep{2011ApJ...743...27T}. The strength of \Cii\ 6580 \AA\ features was found to decrease rapidly with time with nearly no detections by the time of maximum light. However, the number of published spectra available at these early phases is small and it remains unclear if the majority of SNe Ia would show C features if observations were obtained early enough.

Using a new sample of 24 SNe Ia obtained by PTF with at least one spectrum before $-10$ d with respect to maximum, we find that $\sim 40$ per cent of SNe Ia at these early phases have a clear detection of \Cii\ 6580 \AA\ in their spectra. If we include  `absorption?' detections and `flat' profiles, we find that as much as $\sim 55$ per cent may have C present. However, $\sim 25$ per cent of SNe Ia show no clear detection of \Cii\ in their spectra when observed at these early phases.  

Previous studies such as \cite{2012ApJ...745...74F} have shown that the velocity of the \Cii\ 6580 \AA\ feature is lower than expected reaching as low as $\sim$\,11,000 \kms, not much higher than the average \SiII\ 6355 \AA\ velocity. For our sample, the minimum of the \Cii\ 6580 \AA\ feature ranges from $\sim$\,10,700--13,500 \kms. The true range could extend to higher velocities than are measured since at velocities in the range of  $\sim$\,15,000-27,000 \kms, the \Cii\ 6580 \AA\ feature becomes indistinguishable from the much stronger \SiII\ 6355 \AA\ blueward of it \citep{2011ApJ...743...27T,2012ApJ...752L..26P}. 

We estimate that the lowest velocity (measured from the red edge of the absorption profile) at which \Cii\ is present is in the range $\sim$\,7000--11,800 \kms. These relatively low velocities suggest that the unburnt material must be mixed deep into the ejecta and may be caused by macroscopic mixing since spherically symmetric models of deflagration or detonation lead to more burning below 15000 \kms\ \citep{1984ApJ...286..644N,1999ApJS..125..439I,2012ApJ...745...74F}. However, the two-dimensional delayed detonation models of \cite{2010ApJ...712..624M} have C existing as low as $\sim 13000$ \kms, closer to the observed values than the one-dimensional models. Using spectral modelling, \cite{2012ApJ...745...74F} have estimated the mass of C necessary to produce this feature to be 10$^{-3}$--10$^{-2}$ \msun.

We confirm the result of \cite{2011ApJ...743...27T} that SNe Ia where \Cii\ features are identified have narrower light curves than those without a detected \Cii\ 6580 \AA\ feature -- the weighted mean light-curve widths for the `absorption' and `no detection' groups are different at the 3.5-$\sigma$. This could be because more luminous SNe Ia produce more Fe-group elements and have more complete burning and therefore, the amount of \Cii\ remaining is less in more luminous events. Alternatively, in more luminous, higher-stretch events, the \Cii\ 6580 \AA\ feature may be present but at high velocities and therefore, blended with the much stronger \SiII\ 6355 \AA\ feature \citep{2012ApJ...752L..26P}. The study of high-S/N spectra of high-stretch SNe Ia at early times could allow us to investigate the presence of \Cii\ in these objects.

\section{Conclusions}
\label{conclusions}

We have presented the first spectroscopic analysis of the untargeted, low-redshift PTF SN Ia  sample, analysing 359 spectra of 264 SNe at $z<0.2$ at phase up to +5 d with respect to maximum brightness. We have investigated the link between spectral properties and light-curve width, and searched for \Cii\ 6580 \AA\ features and the presence of high-velocity components in the \Caii\ NIR feature, which can be used to distinguish progenitor scenarios. Our main results are as follows.

\begin{enumerate}
\item A high-velocity component is needed to fit the \Caii\ NIR feature in $\sim 95$ per cent of SNe Ia with spectra before $-5$ d with respect to maximum, decreasing to $\sim 80$ per cent around maximum light (Section \ref{pew_phase}, Fig.~\ref{phase}).

\item The velocity of \Caii\ high-velocity component is, on average, significantly larger than the \Caii\ photospheric-component velocity ($\sim 8500$ \kms\ higher) and than the measured \SiII\ velocities (Section \ref{pew_phase}, Fig.~\ref{phase}).

\item SNe Ia falling in the \SiII\ 6355 \AA\ high-velocity ($v>$\,12,000 \kms) subgroup also have higher \Caii\ NIR photospheric-velocity component velocities (Section \ref{normal_high}, Fig.~\ref{col_str}). The lack of a correlation between the \SiII\ 6355 \AA\ velocity and the \Caii\ high-velocity component velocity could suggest that the material producing the \Caii\ high-velocity features is not intrinsic to the SNe, and could be caused by circumstellar material (Fig.~\ref{col_str}).

\item SNe Ia with broader light curve have, on average, a larger contribution from a \Caii\ high-velocity component to their \Caii\ NIR feature relative to a photospheric component \cite[larger $R_{\rm HVF}$, in agreement with the results of][]{2014MNRAS.437..338C}. This is driven mainly by a higher pEW of the \Caii\ high-velocity component in SNe Ia with large $R_{\rm HVF}$ values, but it is also affected by lower pEW of the \Caii\ photospheric-velocity component in these SNe Ia  (Section \ref{hvf_lightcurve}, Fig.~\ref{hvf_stretch}) 

\item SNe Ia with large $R_{\rm HVF}$ values have lower \Caii\ NIR photospheric-velocity component velocities (Section \ref{hvf_specvel}, Fig.~\ref{hvf_vel}). 

\item At least $\sim 40$ per cent (and as much as 55 per cent) of SNe Ia show signs of unburnt material in the form of \Cii\ 6580 \AA\ features if observed at phase earlier than $-$10 d with respect to maximum brightness. Similarly to \cite{2012ApJ...745...74F}, we find that SNe Ia with \Cii\ features tend to have narrower light curves on average (Section \ref{Carbon}).

\item 25 per cent of SNe Ia with a spectrum before $-$10 d with respect to maximum light do not show a \Cii\ 6580 \AA, and these SNe Ia have, on average, broader light curves (Section \ref{Carbon}). This trend may suggest more burning of C in more luminous SNe Ia.

\end{enumerate}

We have confirmed that there is much diversity in the spectral properties of SNe Ia. We have attempted to connect these observed properties and determine if there are distinct subclasses of SNe Ia or trends in their properties. Any explosion models must be capable of explaining the presence of high-velocity \Caii\ features in the vast majority of SNe Ia at early times, the correlation in their strength with light-curve shape, as well as the presence of \Cii\ in at least $\sim 40$ per cent of cases.

\section{Acknowledgements}

K.M. is supported by a Marie Curie Intra-European Fellowship, within the 7th European Community Framework Programme (FP7). M.S. acknowledges support from the Royal Society.  A.G.Y is supported by the EU/FP7-ERC grant no [307260], the Quantum
Universe I-Core program by the Israeli Committee for planning and funding, the ISF,
GIF, Minerva, and ISF grants, and Kimmel and ARCHES awards. N.C. acknowledges support from the Lyon Institute of Origins under grant ANR-10-LABX-66. M.M.K. acknowledges generous support from the Hubble Fellowship and Carnegie-Princeton Fellowship. J.M.S. is supported by an NSF
Astronomy and Astrophysics Postdoctoral Fellowship under award
AST-1302771. A.V.F.'s supernova group at UC Berkeley has received generous financial
assistance from Gary and Cynthia Bengier, the Christopher R. Redlich
Fund, the Richard and Rhoda Goldman Fund, the TABASGO Foundation, and
NSF grant AST-1211916.

The William Herschel Telescope
is operated on the island of La Palma by the Isaac Newton Group in the
Spanish Observatorio del Roque de los Muchachos of the Instituto de
Astrof\'{i}sica de Canarias. Based on observations (GN-2010A-Q-20, GN- 2010B-Q-13, GN-2011A-Q-16 and GS-2009B-Q-11) obtained at the Gemini Observatory, which is operated by the Association of Universities for Research in Astronomy, Inc., under a cooperative agreement with the NSF on behalf of the Gemini partnership: the National Science Foundation (United States), the National Research Council (Canada), CONICYT (Chile), the Australian Research
Council (Australia), Minist\'erio da Ci\^encia, Tecnologia e Inova\c{c}\~ao (Brazil) and Ministerio de Ciencia, Tecnolog\'ia e Innovaci\'on Productiva (Argentina).  Observations were obtained with the
Samuel Oschin Telescope at the Palomar Observatory as part of the
Palomar Transient Factory project, a scientific collaboration between
the California Institute of Technology, Columbia University, La Cumbres
Observatory, the Lawrence Berkeley National Laboratory, the National
Energy Research Scientific Computing Center, the University of Oxford,
and the Weizmann Institute of Science. The authors would like to thank all the observers and data reducers of the collaboration for their hard work throughout the survey (in particular I.~Arcavi, S.~Ben-Ami, J.~Bloom, S.~B.~Cenko, M.~Graham, A.~Horesh, E.~Hsiao, I.~Kleiser, S.~Kulkarni, A.~Miller, E.~Ofek, J.~Parrent, D.~Perley, R.~Quimby, A.~Sternberg, N.~Suzuki, O.~Yaron, D.~Xu). This publication has been made possible by the participation of more than 10,000 volunteers in the Galaxy Zoo: Supernova project (http://supernova.galaxyzoo.org/authors).

The Liverpool Telescope is operated on
the island of La Palma by Liverpool John Moores University in the
Spanish Observatorio del Roque de los Muchachos of the Instituto de
Astrofisica de Canarias with financial support from the UK Science and
Technology Facilities Council.   This work also makes use of observations from the LCOGT network. Some of the data presented herein were obtained at the W.M. Keck Observatory, which is operated as a scientific partnership among the California Institute of Technology, the University of California, and the National Aeronautics and Space Administration (NASA). The Observatory was made possible by the generous financial support of the W.M. Keck Foundation. We thank the dedicated staffs at all the observatories we used for their excellent assistance with the observations. Based on data taken at the European Organisation for Astronomical Research in the Southern Hemisphere, Chile, under program IDs 084.A-0149(A) and 085.A-0777(A). Observations obtained with the SuperNova Integral Field Spectrograph on the University of Hawaii 2.2-m telescope as part of the Nearby Supernova Factory II project, a scientific collaboration between the Centre de Recherche Astronomique de Lyon, Institut de Physique Nucl\'eaire de Lyon, Laboratoire de Physique Nucl\'eaire et des Hautes Energies, Lawrence Berkeley National Laboratory, Yale University, University of Bonn, Max Planck Institute for Astrophysics, Tsinghua Center for Astrophysics, and Centre de Physique des Particules de Marseille.  This research has made use of the NASA/IPAC Extragalactic Database (NED) which is operated by the Jet Propulsion Laboratory, California Institute of Technology, under contract with NASA. 

\bibliographystyle{mn2e}
\bibliography{astro}

\begin{table*}
 \caption{Light-curve information and heliocentric redshift for each SN Ia in the sample. The full table is available online.}
 \label{tab:lightcurve_info}
\begin{tabular}{@{}llccccccccccccccccccccccccccccc}
  \hline
 \hline 
SN& z$_{helio}$&Stretch&MJD of maximum$^2$\\
 \hline
PTF09a&        0.0555$\pm$0.0001		&  0.88$\pm$0.05& 54902.6$\pm$1.0\\ 
PTF09ac&         0.161$\pm$0.001        &  0.97$\pm$0.11& 54917.9$\pm$0.7\\ 
PTF09dhx&        0.087$\pm$0.001        &  1.70$\pm$0.06& 55071.4$\pm$0.4\\ 
PTF09djc&        0.0337$\pm$0.0003      &  0.82$\pm$0.02& 55068.5$\pm$0.1\\ 
PTF09dlc&        0.068$\pm$0.001        &  1.05$\pm$0.02& 55073.8$\pm$0.1\\ 
\hline
\hline
\end{tabular}
\begin{flushleft}
$^1$Heliocentric redshift \\
$^2$Modifed Julian date of maximum $\textit{B}$-band light.\\
\end{flushleft}
\end{table*}

\begin{table*}
 \caption{Spectra measurement information for each spectrum. The uncertainties are only the uncertainties on the fitting routine. The full table is available online.}
 \label{tab:spectral_info}
\begin{tabular}{*{100}{p{1.15cm}}}
  \hline
 \hline 
SN&Phase &v$_{\SiII\ 6355}$&pEW$_{\SiII\ 6355}$&v$_{\SiII\ 4130}$&  pEW$_{\SiII\ 4130}$& v$_{\SiII\ 5972}$& pEW$_{\SiII\ 5972}$& v$_{\Caii\ PVF}$& pEW$_{\Caii\ PVF}$& v$_{\Caii\ HVF}$& pEW$_{\Caii\ PVF}$\\
&(d)&(\kms) &(\AA) &(\kms) &(\AA)&(\kms) &(\AA)&(\kms) &(\AA)&(\kms) &(\AA)\\
 \hline
PTF09a	&	-4.6$\pm$1.0&  9327$\pm$53   & 13.6$\pm$0.8&    9829$\pm$51&    10.8$\pm$0.4  & 10800$\pm$39   &  75.0$\pm$1.0   &   --    & --  &    --    &    --     \\
PTF09ac   &   3.1$\pm$0.7  &  --                   &   --             &  --    &  --  &   10264$\pm$308 &   9.3$\pm$0.7   &   --   &     --    &    --       &     --      \\
PTF09dhx &   -3.4$\pm$0.4 &10826$\pm$301&   23.2$\pm$0.4  & 11209$\pm$307&   28.3$\pm$1.2 &  11351$\pm$299   & 84.9$\pm$0.8 &  11112$\pm$447 &73.4$\pm$2.8 & 15624$\pm$447&      39.3$\pm$1.4 \\
PTF09djc  &  -0.5$\pm$0.1 &11098$\pm$302  &       32.7$\pm$1.5 &  11643$\pm$298&   37.6$\pm$0.3 &  13013$\pm$298&    160.7$\pm$0.5 &  11017$\pm$391&168.9$\pm$45.3 & 15733$\pm$391&   123.4$\pm$44.1 \\
PTF09dlc  & -10.8$\pm$0.1 &12376$\pm$318 &    3.8$\pm$0.5      &  --   &  --& 16947$\pm$300&    96.9$\pm$1.0       &    --   &    --      &     --    \\
\hline
\hline
\end{tabular}
\begin{flushleft}
$^1$Heliocentric redshift \\
$^2$Modifed Julian date of maximum $\textit{B}$-band light.\\
\end{flushleft}
\end{table*}

\end{document}